\documentclass[useAMS,usenatbib]{mn2e}

\usepackage{graphicx}
\usepackage{multirow} 
\usepackage{pdflscape}
\usepackage{amssymb}
\def\lesssim{\la}

\newcommand{\aap}{A\&A}

\newcommand{\apj}{ApJ}
\newcommand{\apjs}{ApJS}
\newcommand{\aj}{AJ}
\newcommand{\apjl}{ApJL}
\newcommand{\araa}{ARA\&A}
\newcommand{\mnras}{MNRAS}

\newcommand{\pasp}{PASP}

\begin{document}
\title[Dippers and Dusty Disk Edges]{Dippers and Dusty Disk Edges: A Unified Model}
\author[Bodman et al.]{Eva H. L. Bodman$^{1,2}$, Alice C. Quillen$^1$, Megan Ansdell$^3$, Michael Hippke$^4$,   
\newauthor 
Tabetha S. Boyajian$^{5,6}$, Eric E. Mamajek$^{7,1}$, Eric G. Blackman$^1$,  Aaron Rizzuto$^8$, 
\newauthor 
and Joel H. Kastner$^9$  \\
$^1$Department of Physics and Astronomy, University of Rochester, Rochester, NY 14627, USA \\
$^2$ NASA Postdoctoral Program, Nexus for Exoplanet System Science \\
$^3$Institute for Astronomy, University of Hawai'i at M\={a}noa, Honolulu, HI 96822, USA \\
$^4$Institute for Data Analysis, Luiter Stra{\ss}e 21b, 47506 Neukirchen-Vluyn, Germany\\
$^5$Department of Astronomy, Yale University, New Haven, CT 06511, USA\\
$^6$Department of Physics and Astronomy, Louisiana State University, Baton Rouge, LA 70803 USA \\
$^7$Jet Propulsion Laboratory, California Institute of Technology, M/S 321-100, 4800 Oak Grove Dr., Pasadena, CA 91109, USA \\
$^8$Department of Astronomy, The University of Texas at Austin, Austin, TX 78712, USA\\
$^9$Chester F. Carlson Center for Imaging Science, Rochester Institute of Technology, Rochester, NY 14623-5603, USA\\
%List and order to be determined.
}
%\date{}                                           % Activate to display a given date or no date

\maketitle
\begin{abstract}
We revisit the nature of large dips in flux from extinction by dusty circumstellar material that is observed by \textit{Kepler} for many young stars in the Upper Sco and $\rho$ Oph star formation regions. These young, low-mass ``dipper" stars are known to have low accretion rates and primarily host
%A search for dips in observed stellar flux, ``dippers", in the Upper Scorpius and $\rho$  Ophiuchus  star formation regions with the Kepler mission by \cite{2016A} identified young, low mass stars with low accretion rates and primarily hosting 
moderately evolved dusty circumstellar disks. Young low mass stars often exhibit rotating star spots that cause quasi-periodic photometric variations. We found no evidence for periods associated with the dips that are different from the star spot rotation period in spectrograms constructed from the light curves. The material causing the dips in most of these light curves must be approximately corotating with the star.  We find that disk temperatures computed at the disk corotation radius are cool enough that dust should not sublime. 
%If material needs to be cooler than the dust sublimation temperature, then dippers are preferentially associated with young, low mass stars which is consistent with the sample. 
Crude estimates for stellar magnetic field strengths and accretion rates are consistent with magnetospheric truncation near the corotation radius. Magnetospheric truncation models can explain why the dips are associated with material near corotation and how dusty material is lifted out of the midplane to obscure the star which would account for the large fraction of young low mass stars that are dippers. We propose that variations in disk orientation angle, stellar magnetic field dipole tilt axis, and disk accretion rate are underlying parameters accounting for differences in the dipper light curves. 

\end{abstract}
\begin{keywords}
stars: variables: T Tauri -- accretion, accretion discs --
\end{keywords}

\section{Introduction}

Optical and infrared light curves of pre-main sequence stars display variability and this has been interpreted to be due to a variety of mechanisms, including rotation of hot and cold star spots, accretion hotspots, accretion bursts and occultations by dusty disk structures such as warps, clumps, vortices  and accretion streams (e.g., \citealt{bouvier03, plavchan08, 2010Alencar, 2014Cody}). Recent photometric surveys have identified a class of young stars, known as ``dippers'', whose light curves are relatively flat but also show deep (10-50\%) quasi-periodic or aperiodic drops in flux lasting up to a few days \citep{2010Alencar, moralescalderon11, parks14, 2014Cody, stauffer15, 2016A}.

As the dips are too deep to be caused by star spots, most scenarios consider variable levels of extinction to account for the dipper phenomenon. \citet{moralescalderon11} found that the wavelength dependence of the dips is roughly consistent with that expected for a standard interstellar extinction law, as previously reported for AA Tau \citep{bouvier03}. However, \citet{stauffer15} saw similar depth in 4.5$\mu$m and I band photometry for Mon~1165 and so inferred that the occulting dusty material could be optically thick. In some stars, optical and infrared flux variations are anti-correlated suggesting that the same material causing occultation in a disk edge is also illuminated by the star and reradiates the energy in the infrared \citep{2015McGinnis}. 

Among classical T~Tauri stars (CTTS) in NGC~2264, approximately 20\% to 30\% are dippers \citep{2010Alencar,2014Cody}. The large fraction of young stars exhibiting dipper behaviour indicates that it is a ubiquitous phenomenon. However, circumstellar disks are thin.  With disk aspect ratio (scale height over radius) estimated to be approximately 0.1, dusty material must be lifted high above the midplane to account for the large fraction of randomly oriented stars in clusters that are dippers. As voiced by \citet{stauffer15}, ``A general concern with any model for the variable extinction stars is simply that there seems to be too many of them in NGC~2264.''

During the second campaign (C2) of the Kepler Observatory's K2 mission \citep{howell14}, the star formation regions in Upper Scorpius (Upper Sco) and $\rho$ Ophiuchus ($\rho$ Oph) were observed. The typical ages of the stars in the Upper Sco and $\rho$ Oph regions are $\sim10$ Myr \citep{2012Pecaut} and $\sim1$ Myr \citep{2007Andrews}, respectively. 
 \cite{2016A} searched approximately 13,000 of the K2/C2 light curves for dippers and generated a sample of 25 confirmed dippers; 10 of these they presented with follow-up spectroscopic observations and analyses of archival data. All of the 10 well-characterised dipper stars found were late-K or M dwarf stars, despite the entire K2/C2 sample of the star forming regions containing little bias towards late-type stars above that expected from the initial mass function. \cite{2016A} estimate a $\sim 0.1\%$ chance of randomly choosing only late-type stars out of the K2/C2 sample and notes that $\sim 80\%$ of previously identified early type and $\sim 95\%$ of late type young stars in these two star formation regions were observed by K2/C2. A reexamination of the sample of dippers studied by \citet{stauffer15} in NGC~2264 (see their Table 4) confirms that most known dippers are  late-type stars. \citet{2016A} also presented infrared spectral energy distributions, finding that the Upper Sco and $\rho$ Oph dippers host moderately-evolved protoplanetary disks (pre-transition disks) with evidence of depressions or clearings in their dusty disks. Emission line equivalent widths in H$\alpha$ and Pa$\gamma$ presented in their Table 3 show that accretion rates are low, consistent with the later stage of disk evolution implied by the infrared spectral energy distributions. 
 This preference for moderately-evolved disks is consistent with the fact the dippers in NGC~2264 also host more evolved disks \citep{sousa16, 2015Venuti}. If the dipper phenomenon is related to dust extinction from structures in the inner disk \citep[e.g.][]{2016A, 2015McGinnis, bouvier03}, then the tendency for the dipper phenomenon to be associated with later-type stars may reflect the fact that such stars tend to have longer primordial disk lifetimes \citep{2012Luhman}. In the Upper Sco region, K-type stars have a disk fraction of $\sim10\%$ whereas $\sim25\%$ of late-M-type stars host disks \citep{2012Luhman}.

In this study, we will attempt to differentiate between the various proposed mechanisms for causing dippers by reanalysing the sample compiled by \citep{2016A}. %Previous observational studies of the dippers in NGC~2264 searched a known sample of CTTS \citep[e.g.,][]{2015McGinnis, stauffer15}. We investigate a sample created by \citet{2016A} from a search of all K2/C2 objects. 
We expand upon the characterisation of the dipper sample by \citet{2016A} and reanalyse the light curves to more thoroughly examine the possible mechanism for the phenomenon.% going beyond Andsell to test the mechanism models and in a way different from McGinnis. 
 Using spectrograms, we study the periodicity of the light curves of the stars in the \citet{2016A} sample and search for evidence of multiple or changing periods. We also characterise the shape of the dips to constrain the location of the obscuring material in the disk for aperiodic dippers as well as periodic. Finally, we propose a model that explains why low-mass, low-accretion rate stars are preferentially found in a search for dippers and invokes a stellar occultation geometry involving dust far out of the midplane, thereby accounting for the large fraction of low-mass stars displaying the phenomenon and the variety of dipper light curve morphologies. 

\section{Upper Sco and $\rho$ Oph Dippers}\label{sec:sample}

Table \ref{tab:starp} lists the 25 dipper stars discovered by \citet{2016A}. The first column in Table 1 lists the K2 Ecliptic Plane Input Catalog (EPIC) ID, the second column is the 2MASS Point Source Catalog identifier and the third column the J band magnitude from the 2MASS Point Source Catalog \citep{cutri03}. Table \ref{tab:starp} begins with the 10 stars \citet{2016A} studied in detail by obtaining new and archival multi-wavelength data. The remaining 15 stars, which were found while studying the first ten, are listed only by EPIC number in Table 5 by \citet{2016A} and for them we fill in the Table with measurements from the literature.  

 For the first ten stars, we list (in columns  4-7) in order cluster membership, rotation period ($P$), spectral type and stellar temperature ($T_\mathrm{eff}$) estimated from the spectra reported by \cite{2016A} (see their Tables 1 and 3).
Using the extinction ($A_V$) from their Table 3, a distance of 140 pc for both clusters \citep{2008Mamajek}, bolometric corrections from \cite{2013Pecaut} and the 2MASS J band magnitudes, we compute the bolometric luminosities ($L_\mathrm{Bol}$) and these are listed in column 8 in our Table \ref{tab:starp}.

Two stars in Table \ref{tab:starp} lack effective temperature or spectral type measurements in the literature and three stars have spectral type measurements but lack extinction estimates.  We have left blank the parts of the table that depend on stellar luminosity for these 5 stars.  We estimate the rotation period $P$ by identifying a dominant peak in the autocorrelation function of the normalised light curve and by checking for a consistent dominant peak at the same frequency in the periodogram  (computed using python's routine \texttt{scipy.signal.periodogram} 
with Bartlett's method \citep{1950Bartlett}). This period is typically associated with sinusoidal variations in the light curve from star spots but could be associated with the dips instead.
 For the additional fifteen stars, we list spectral types, bolometric luminosities and cluster memberships reported in the literature, when available. We estimate the temperature from the spectral type using conversions from \cite{2013Pecaut}.% (their Table 5). 

\begin{landscape}
\begin{table}
\begin{center}
\vbox to 125mm{\vfil
\caption{\large Stellar Properties}
\begin{tabular}{@{}llcccccccccccc}
EPIC ID  	   & 2MASS ID      		&  J   	& Mem.   	  & $P$ 		& SpT		& $T_\mathrm{eff}$	& $L_\mathrm{Bol}$	& $R_\star$ 	& $M_\star$	& Log(age)	& $T_\mathrm{cor}$	& $R_\mathrm{cor}$	& Var.  \\
                	   &                        		&   mag 	&              	  & day		& 			&    K             		&    $L_\odot$        	& $R_\odot$	&   $M_\odot$ 	& Log(year)	&     K                       	& $R_\odot$		& 		\\
	  (1)     &     (2)                   	&       (3)   &        (4)    &    (5)    		& (6)			&     (7)                      &   (8)                	&  (9)                &  (10)         	&  (11)              	&         (12)       		&   (13)    			& (14) 	\\
\hline
203343161 & 16245587-2627181 	& 12.172 	& USc	 & 2.24   		& M5.5$^a$	& $3120^a$		&   0.037       		&  0.66   		& 0.32      	       	& 7.25		& 810                       	&  4.9               	 	&  Q  	\\
203410665 & 16253849-2613540 	&  8.688  	& Oph	 & 4.24  		& K7$^a$ 		& $4050^a$ 		&  1.5           		& 2.5  		& 0.75      	        	& 6.0			& 1430                     	&   10.0        		&  A  	\\
203895983 & 16041893-2430392 	&  9.975  	& USc	 & 2.44   		& M2.5$^a$	& $3655^a$		&   0.35        		& 1.5    		& 0.55      	       	& 6.4 		& 1250                      	&  6.3            		& A   	\\
203937317 & 16261706-2420216 	&  9.655  	& Oph	 & 5.44   		& K7.5$^a$	& $4070^a$		&  0.95          		& 2.0  		& 0.7               	& 6.1 		& 1190                      	&  11.6         		&  Q  	\\
204137184 & 16020517-2331070 	& 11.730 	& USc	 & 2.64   		& M4$^a$ 	& $3210^a$		& 0.062          	 	& 0.81 		& 0.4        	        	& 7.1 		&  840                       	&  5.9           		& Q   	\\
204630363 & 16100501-2132318 	& 10.069 	& USc	 & 6.66   		& K7.5$^a$	& $4070^a$ 		&  0.41         	 	& 1.3    		& 0.7      	       	& 6.6 		&  900                       	& 13.2          		& A   	\\
204757338 & 16072747-2059442 	& 11.229 	& USc	 & 2.39   		& M4.5$^a$	& $3145^a$		&  0.11        	  	&1.1   		& 0.35            	& 6.6 		&  1030                     	&   5.3          		& Q   	\\
204932990 & 16115091-2012098 	& 11.452 	& USc	 & 2.30   		& M3.5$^a$	& $3390^a$		&  0.096         	 	& 0.90    		& 0.55      	       	& 7.15		& 930                       	&   6.0          		& A   	\\
205151387 & 16090075-1908526 	& 10.220 	& USc	 & 9.55     		& M1$^a$ 	& $3975^a$		&  0.35         	 	& 1.2   		& 0.7                	& 6.55		& 760                       	&   16.8        		& Q   	\\
205519771 & 16071403-1702425 	& 11.754 	& USc	 & 2.46   		& M3.5$^a$	& $3390^a$		&   0.071       		& 0.77    		& 0.5        	        	& 7.3 		& 850                        	&    6.1         		& A   	\\
\hline
203429083 & 15570350-2610081 	& 13.998 	&            	  & 1.56 	    	&			&                           	&                           	&                     	&                  	&                     	&                      		&       	 	    	& A   	\\
203824153 & 16285407-2447442 	& 10.683 	& Oph      	  & 11.59	    	& M2$^e$		& 3490	          	& 0.62$^b$         	& 2.2 		& 0.45              	& 6.0	      		& 880                       	& 16.5    		    	& Q   	\\
203843911 & 16262367-2443138 	&  9.391  	& Oph      	  & 8.96	    	& K5$^e$ 		& 4140	          	& 1.40$^b$         	& 2.3 		& 0.75             	& 6.0 		& 1090                     	& 16.5     	 	    	& Q   	\\
203850058 & 16270659-2441488 	& 12.433 	& Oph      	  & 2.88  	    	& M5$^f$ 		& 2880	          	& 0.10$^b$         	& 1.3  		& 0.1             	& 6.15 		& 1140                     	& 4.0      		    	& Q    	\\
203862309 & 16274270-2438506 	& 13.245 	& Oph      	  & 4.47   	    	& M2$^g$		& 3490	          	& 0.21$^b$         	& 1.3 		& 0.45             	& 6.5 		& 930                       	& 8.8       		    	& Q    	\\
203969672 & 16270907-2412007 	& 12.410 	& Oph      	  & 5.03   	    	& M2.5$^g$	& 3430	          	& 0.39$^b$         	& 1.8  		& 0.35             	& 6.0 		&1090                      	& 8.7       		    	& A   	\\
203995761 & 16281673-2405142 	& 10.984 	& Oph      	  & 5.26        	& K5$^h$		& 4140	         	& 0.72$^b$         	& 1.6 		& 0.73              	& 6.3 		& 1110                     	& 11.5     	    		& A   	\\
204107757 & 15560104-2338081 	& 13.856 	& USc      	  & 1.52        	& M6.5$^i$	& $<$2880               	&                           	&                     	&                  	&                      	&                      		&       	    		& A   	\\
204211116 & 16214199-2313432 	& 12.232 	& Oph      	  & 2.14        	& M2$^j$ 		& 3490	    	 	& 0.070$^b$       	& 0.7 		& 0.6           	& 7.45 		& 860                      	& 5.9       	    		& Q   	\\
204329690 & 16220194-2245410 	& 12.230 	&  	 	  & 2.18        	&			&                         	&                             	&                   	&                   	&                     	&                    		&       	    		& A   	\\
204449274 & 16222160-2217307 	& 13.742 	& USc      	  & 1.56         	& M5$^c$		& 2880      		& 0.01$^c$            	& 0.5 		& 0.18             	& 7.45		& 720                       	& 3.2            		& Q   	\\
204489514 & 16030161-2207523 	& 12.731 	& USc      	  & 1.79         	& M4.75$^k$	& 2950                	&                          	&                     	&                   	&                     	&              	        		&      	 	   		& A   	\\
204530046 & 16105011-2157481 	&  11.863 	&             	  & 2.06       	&			&                         	&                           	&                     	&                   	&                     	&              		        	&       			&  A  	\\
204864076 & 16035767-2031055 	&   9.608 	& USc      	  & 3.95         	& K5$^l$ 		& 4140             		&                          	&                     	&                   	&                      	&            	          	&       			&  A  	\\
205068630 & 16111095-1933320  	& 12.338 	& USc      	  & 2.08         	& M5$^d$		& 2880		        	& 0.03$^d$        	& 0.8 		& 0.2     	        	& 6.95 		& 880                       	& 4.0       			&  A  	\\
\hline
\end{tabular}
{\\ Columns:(1) EPIC Identifier, (2) 2MASS Identifier, (3) J magnitude from the 2MASS Point Source Catalog \citep{cutri03}, (4) cluster membership, (5) rotation period, (6) spectral type, (7) effective temperature from \cite{2016A} or estimated from spectral type using \cite{2013Pecaut}, (8) bolometric luminosity, (9) stellar radius, (10-11) stellar masses and ages estimated from pre-main-sequence models by \citet{2014Chen}, (12) temperature at corotation radius from equation \ref{eqn:temp}, (13) corotation radius from equation \ref{eqn:rcor}, (14) variability type: A-Aperiodic, Q- Quasi-periodic \\
Refs: $^a$\cite{2016A}, $^b$\cite{2009Evans}, $^c$\cite{2008Slesnick}, $^d$\cite{2002Preibisch}, $^e$\cite{1995Kenyon}, $^f$\cite{2015Manara}, $^g$\cite{2005Wilking}, $^h$\cite{1998Martin}, $^i$\cite{2004Martin}, $^j$\cite{2011Gully}, $^k$\cite{2012Luhman}, $^l$\cite{1999Preibisch} \\
\label{tab:starp}}
\vfil} 
\end{center}
\end{table}
\end{landscape}

We used the stellar effective temperatures and bolometric luminosities (listed in Table \ref{tab:starp} in columns 5 and 6) to calculate the radius of each star, $R_\star$, in solar units and this is listed in the Table \ref{tab:starp} under column 9. 
The effective temperature are known to about $\pm200$ K and $\log(L_\mathrm{Bol})$ to $\pm0.1$ so the error for the stellar radius is of order 10\%.
To estimate the stellar mass, $M_\star$, we placed the effective temperature and bolometric luminosity on PARSEC isochrones \citep{2014Chen} using solar metallicity. Since stellar ages can vary within the star formation regions, we estimate the age of each star along with its mass by matching both the temperature and the luminosity to the pre-main sequence  tracks.  The estimated masses and ages are listed in Table \ref{tab:starp} under columns 10 and 11. The estimated ages are consistent with the ages for Upper Sco and $\rho$ Oph  of $\sim10$ Myr \citep{2012Pecaut} and $\sim1$ Myr \citep{2007Andrews}, respectively.
With large systematic uncertainties in isochrones \citep{2004Hillenbrand} and the uncertainties in the effective temperature and luminosity, the uncertainties in age and masses can be as large as a factor of $\sim2$.

Following the definitions used by \cite{2015McGinnis}, a quasi-periodic dipper has dips that occur at approximately-periodic intervals but with depths and shapes that change from dip to dip whereas aperiodic dippers have dips that are stochastically distributed. In Table \ref{tab:starp} we list the variability type Q for quasi-periodic or A for aperiodic for the first ten as reported by \citet{2016A}.   
The remaining 15 stars we classified as Q or A ourselves by looking at phase-folded light curves, folded using the periods listed in Table \ref{tab:starp}.
The light curves, along with spectrograms for the dippers listed in Table \ref{tab:starp}, are shown in Figure \ref{fig:SLC1}.
We used the K2/C2 K2SFF light curve data released to MAST that was corrected with the Self Field Flattening technique \citep{2014Vanderburg}. The light curves are normalised by dividing by the median measured brightness.

\begin{figure*}
\includegraphics[width=6.6in, trim= 0 0 0 0 ]{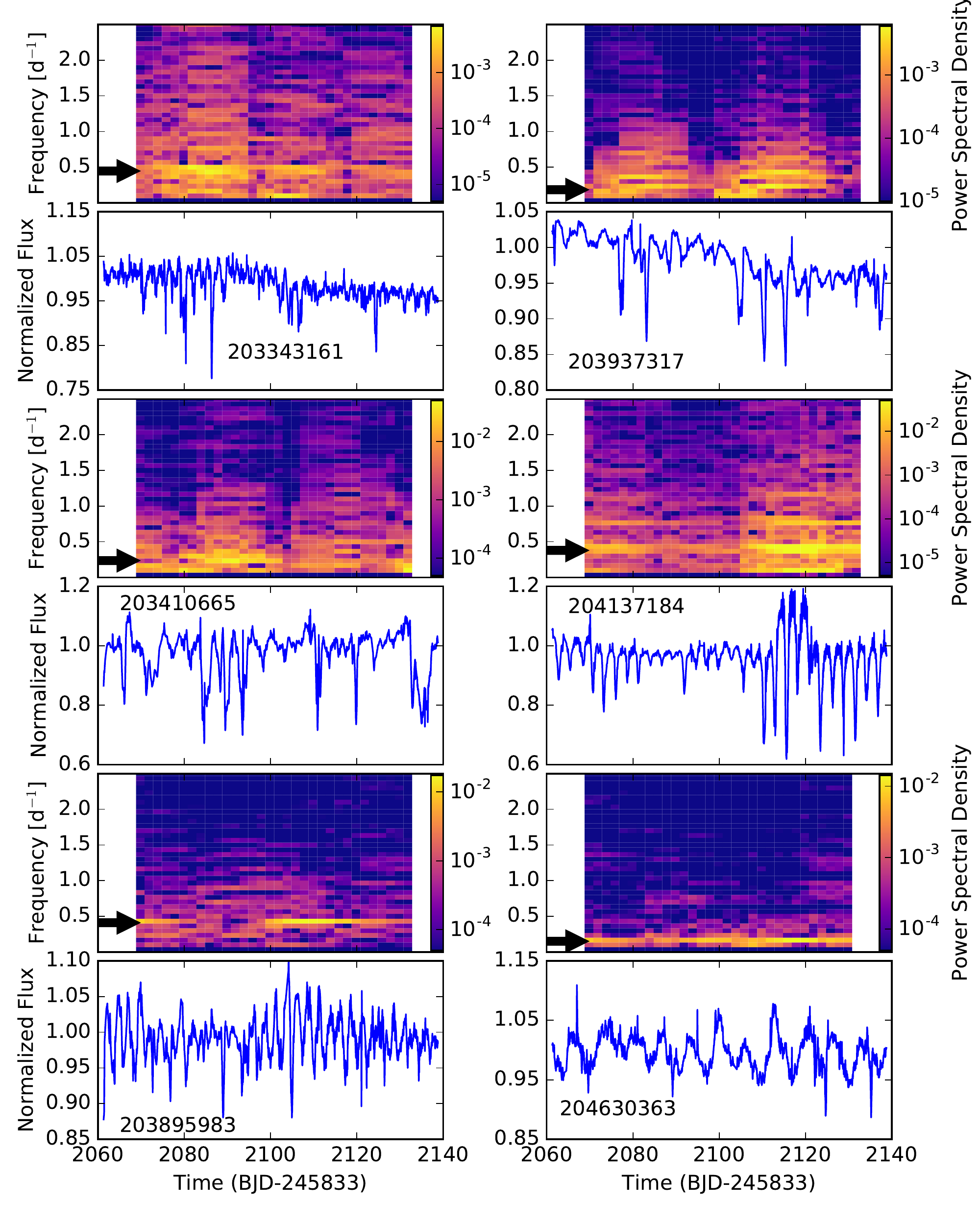}
\caption{Light curves of dippers with the associated spectrogram plot above the light curve plot. 
For the spectrograms and the light curves, $x$ axes are time (Barycentric Julian Date) and over the same interval.
  Each light curve is labeled by the star's EPIC ID and they are shown in the order listed in Table \ref{tab:starp}. In the spectrograms the colour shows the log of the spectral power density and ranges from purple (low power) to yellow (high power). {The arrow on each spectrogram marks the period for that star listed in Table \ref{tab:starp}.}
}
\label{fig:SLC1}
\end{figure*}

\setcounter{figure}{0}

\begin{figure*}
\includegraphics[width=6.6in, trim= 0 0 0 0 ]{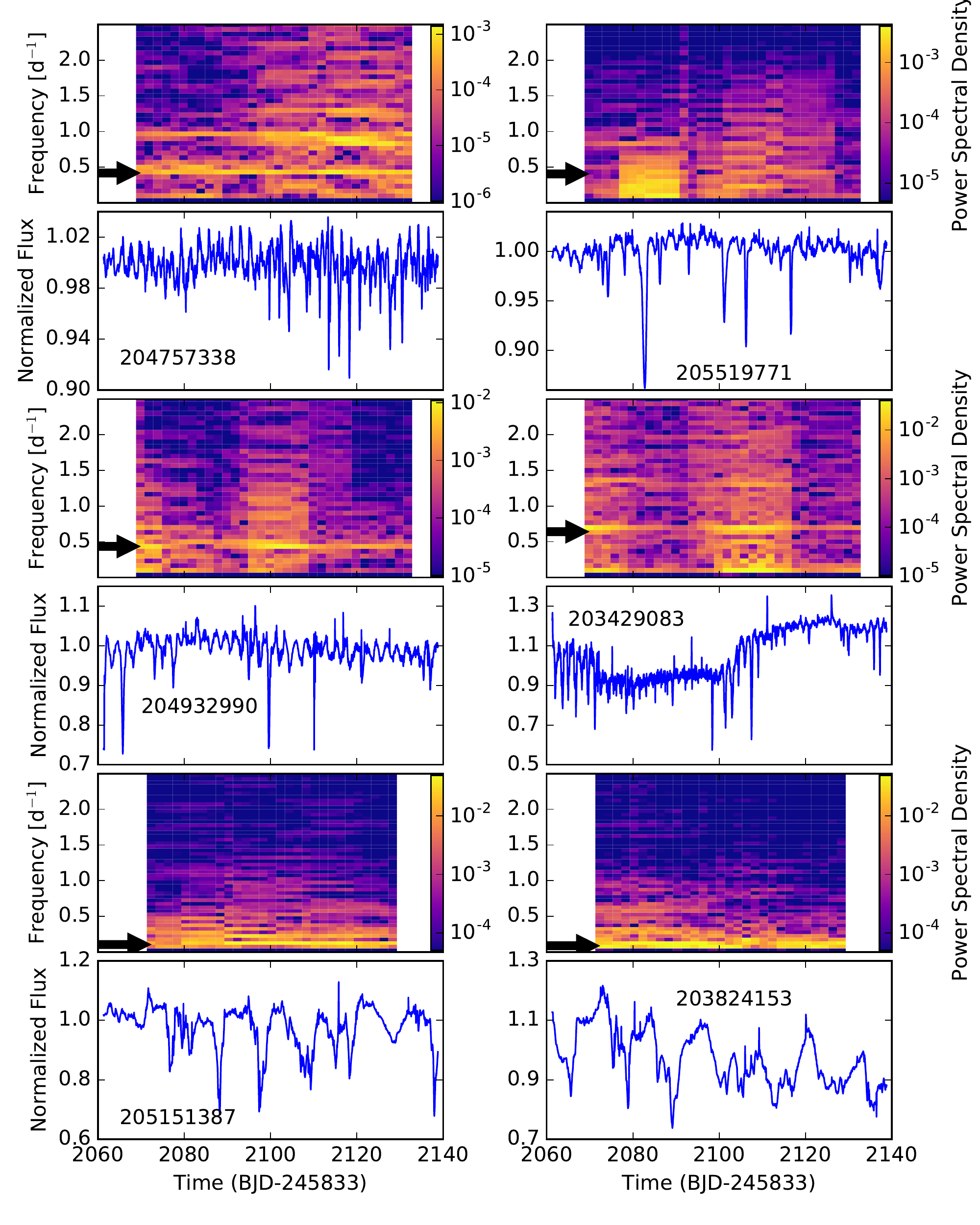}
\caption{continued.
}
%\label{fig:SLC2}
\end{figure*}
\setcounter{figure}{0}

\begin{figure*}
\includegraphics[width=6.6in, trim= 0 0 0 0 ]{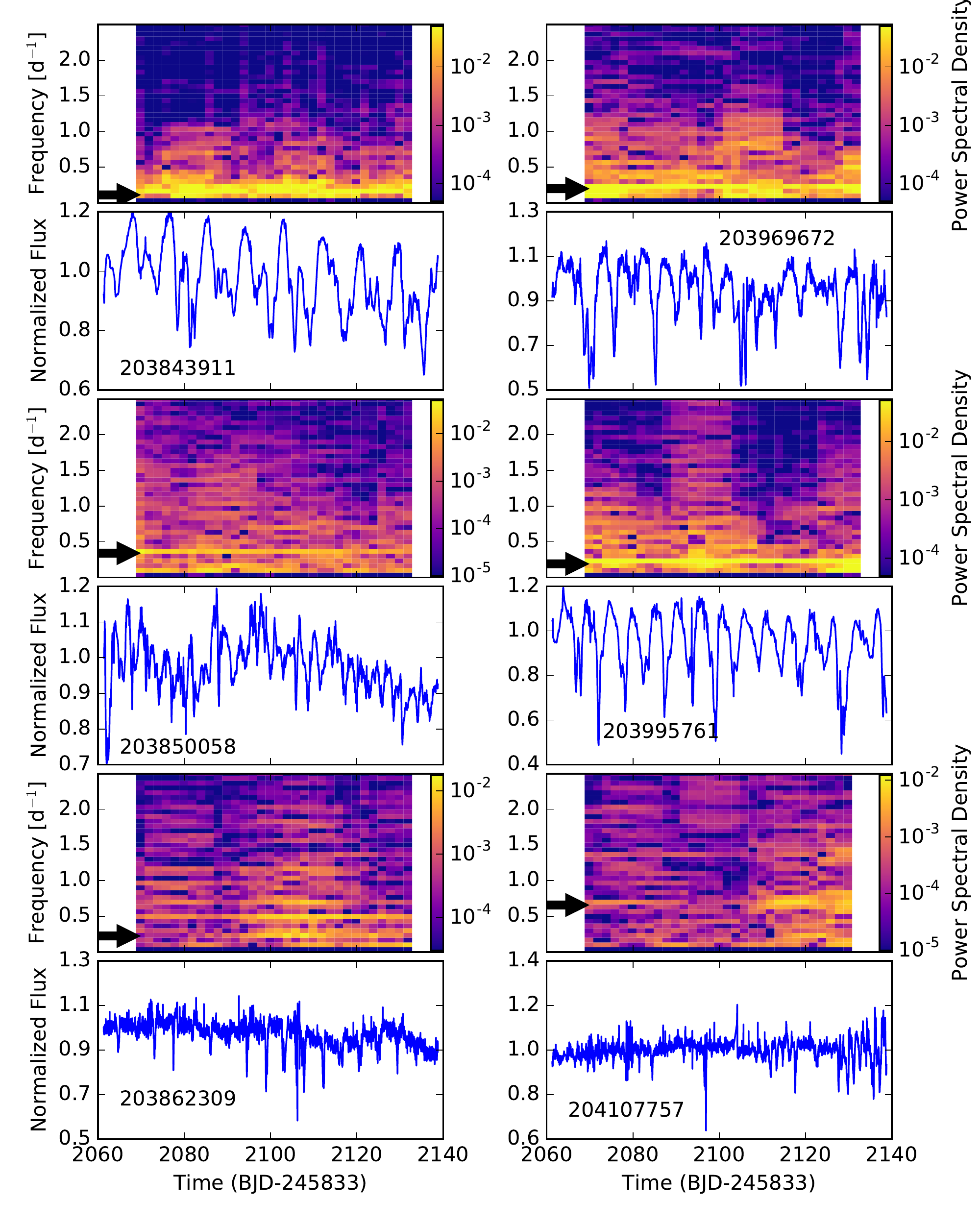}
\caption{continued.
}
%\label{fig:SLC3}
\end{figure*}
\setcounter{figure}{0}

\begin{figure*}
\includegraphics[width=6.6in, trim= 0 0 0 0 ]{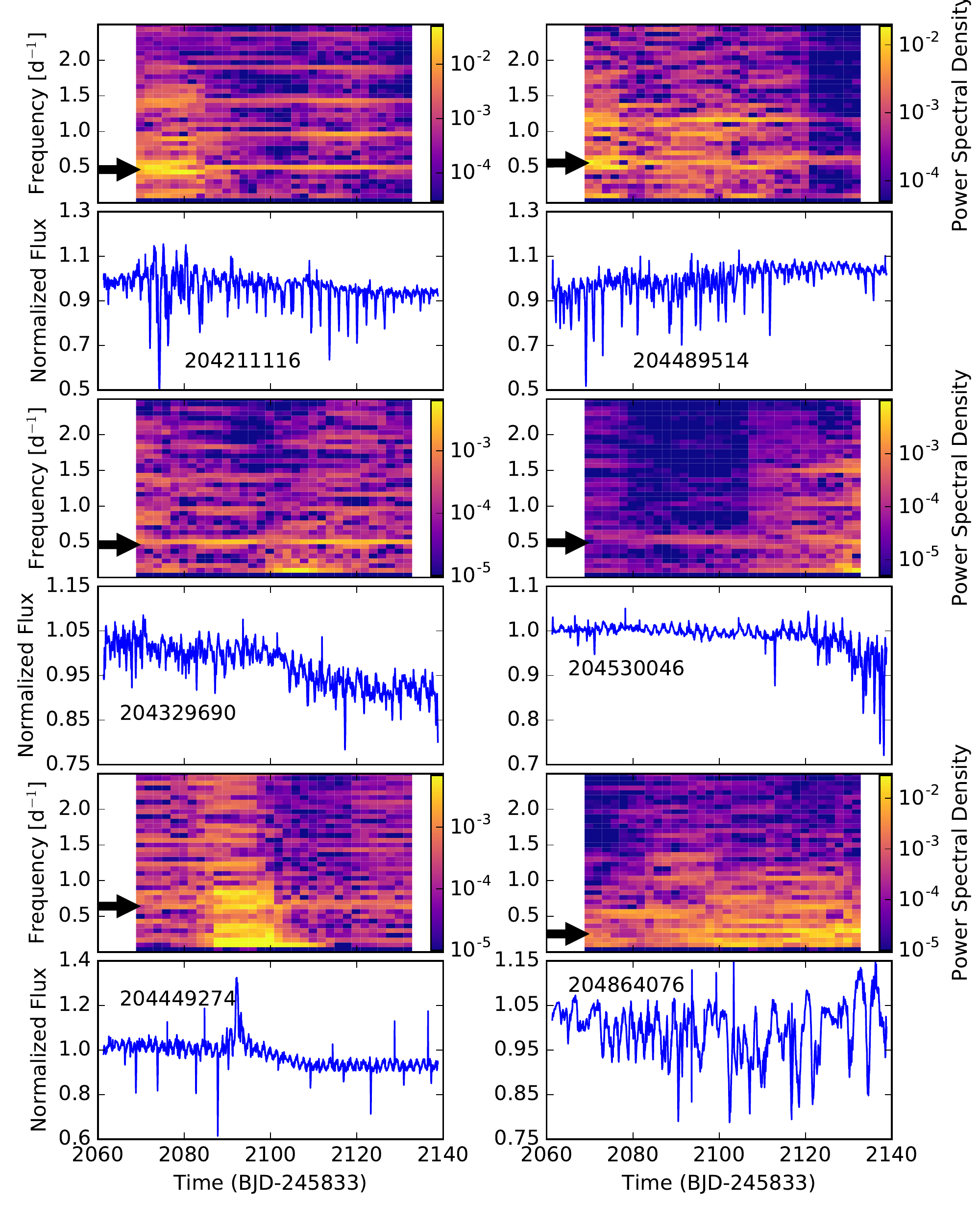}
\caption{continued.
}
%\label{fig:SLC4}
\end{figure*}
\setcounter{figure}{0}

\begin{figure}
\includegraphics[width=3.3in, trim= 0 0 0 0 ]{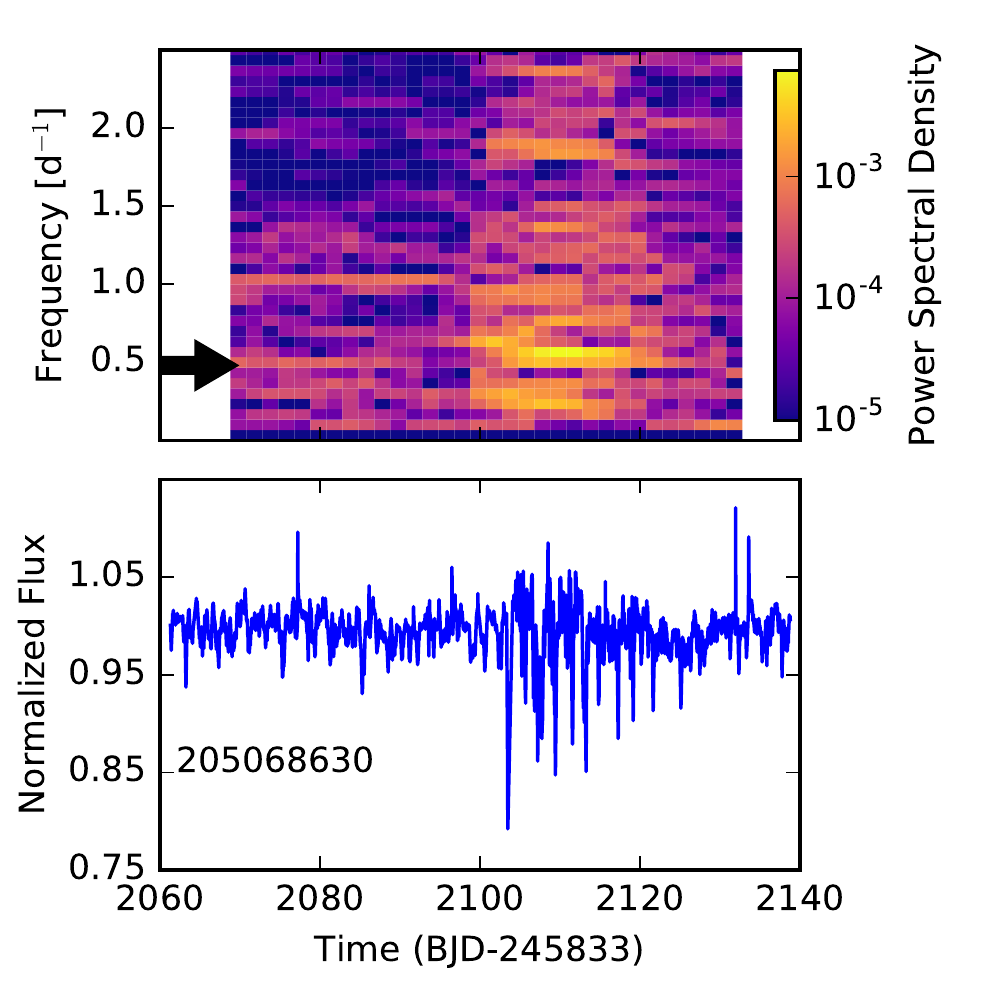}
\caption{continued.
}
%\label{fig:SLC5}
\end{figure}

\subsection{Star vs Disk rotation and Spectrograms}\label{sec:spec}

As pre-main sequence stars rotate, %(e.g., Figure 1 by \citealt{gallet15}, Figure 4 by \citealt{2010Alencar}), 
their light curves can exhibit sinusoidal variations from star spots (e.g., \citealt{herbst06}).  Periods for pre-main sequence stars with mass near $0.4 M_\odot$ are typically a few days, and so are similar to those listed in Table \ref{tab:starp} (e.g., see histograms in Figure 2 by \citealt{herbst06} and the left panel of Figure 1 by \citealt{gallet15}). \citet{2016A} referred to the periods measured from the dipper light curves as rotation periods, implying that they were associated with stellar rotation. However, many of the dips in the light curves are too deep to be associated with star spots. If the obscuration is due to dust, it must be located distant enough from the star that the dust is below dust sublimation temperature, $T_\mathrm{sub} \sim 1200 -1600K$. The exact sublimation temperature depends on the species of dust such as olivine ($T_\mathrm{sub} \approx 1300$ K) or pyroxene ($\approx1600$ K) \citep{2011Kobayashi}. The dipper absorption is more plausibly associated with obscuration from more distant colder disk material rather than structures on the stellar photosphere so the measured periods are not necessarily equal to the rotation period of the star.

\citet{stauffer15} looked for evidence of separate dip and stellar rotation periods in the young stars in NGC~2264 but did not see multiple oscillation frequencies  in the light curves. However for at least 3 stars (Mon-21, Mon-56, Mon-378, and Mon-1580), the authors assigned semi-sinusoidal variations to photospheric star-spots and the short- duration flux dips, with the same period, to dust in or near the co-rotation radius in a circumstellar disk. They inferred that the disk was locked to the stellar rotation, in harmony with expectations from models of angular momentum evolution of young accreting stars \citep{konigl91,collier93}. The same conclusion was reached by \citet{2007Bouvier,2015McGinnis}.
 
We looked for evidence of multiple periods in our sample by computing spectrograms for all 25 light curves. Prior to computing the spectrograms we applied a median filter to remove extreme outliers and we filled in gaps in the light curve with median values so that they were evenly sampled (in time). Light curves for three stars (EPIC~203937317, 204107757, 204137184) have artefacts which we removed by deleting the corrupted data points and filling the gaps with a normalised flux of 1.0. We computed spectrograms with the python routine \texttt{scipy.signal.spectrogram} which uses a discrete Fourier transform that requires evenly sampled data \citep{1999Oppenheim}. To compute each spectrum, we used a window function (in time) that is a Gaussian function multiplied by a top-hat function (1 inside and zero outside).  The Gaussian is centred at the middle of the top-hat. For stars with periods shorter than 9.5 days the top-hat is $\sim15$ days wide and the Gaussian function has a standard deviation of 12.0 days. For stars with longer periods the window is 20 days wide and the Gaussian has a standard deviation of $\sim15.5$ days. In both cases, we moved the center of the window in steps of two days to compute each spectrum. The resulting 2-dimensional spectrogram  is shown as a 2-dimensional image along with the light curves in Figures \ref{fig:SLC1}. The colour of each spectrogram shows the log of the spectral power density.

Inspection of our spectrograms reveals a single dominant frequency in most of the light curves and for all but one star (EPIC~204864076, discussed below), this period is associated with the dips in the light curve. 
The dominant frequency in the spectrograms agrees with the period listed in Table \ref{tab:starp} which is marked in each spectrogram with an arrow. 
When the dips are deep, harmonics of that frequency are usually present (for example see EPIC~204211116's spectrogram). We don't see a different frequency appear in the spectrograms when the dipper phenomenon is absent. For example, the later half of the light curve of EPIC~204489514 does not display dippers but the period of oscillations is the same as that in the earlier part of the light curve. 
Six stars, EPIC~204449274, 204489514, 204530046, 204757338, 204932990 and 205519771, display clear sinusoidal variations that are probably due to star spots in addition to a mix of quasi-periodic and aperiodic dips. 
Two more stars, EPIC~203937317, 204630363 also show sinusoidal variations in their phase-folded light curves that could be from starspots. In some cases, dips appear at a similar phase, but can be separated by more than one rotation period.  For example, the dips in EPIC~204449274 are separated by times larger than the stellar rotation period but occur at the same phase whereas the dips in EPIC 205519771, while also separated by multiple rotation periods, occur at different phase and are therefore aperiodic.

Only one star does not exhibit a single dominant frequency. EPIC~204864076 displays oscillations at a period of about 2 days in a 20 day region at the beginning of the light curve followed by a series of aperiodic dips. The strongest peak in the periodogram has a period of about 4 days and that is what we have listed in Table \ref{tab:starp}, even though it may not be from  stellar rotation.

In addition to the dips, we probably see periodic variability associated with star spots.  However, in most cases, the period associated with starspots is so similar to that associated with the deep dips that we cannot measure a difference between them. This suggests that the inner disk is locked to stellar rotation, supporting the finding by \citet{stauffer15} for dipper stars in NGC~2264. 

 AA~Tau showed the same quasi-periodic behaviour at approximately the same period over 20 years of observations \citep{bouvier13}. Similarly, \citet{2015McGinnis} found little difference in periods measured in AA~Tau-like stars when comparing light curves from 2008 and 2011 
 but about half of their sample switched between aperiodic and AA~Tau-like behaviour. This change between quasi-periodic and aperiodic behaviour is further discussed in section \ref{sec:magacc}.
  Even though the K2/C2 light curves shown here only extend about 80 days, they confirm trends found from these previous studies because we see no significant change in the dominant periods in the spectrograms, except for EPIC~204864076.

\subsection{Disk Temperature at Corotation} \label{sec:temp}

If the dips in the light curve are due to absorption of star light from dusty material then that material must have a temperature below the dust sublimation temperature. The period evident in the light curves is associated with a corotation radius, $R_\mathrm{cor}$, where gas in a circumstellar disk orbits with the same period as the star. Using the rotation periods and stellar masses listed in Table \ref{tab:starp}, we estimate with Kepler's third law the radius for corotating material 
\begin{equation}
R_\mathrm{cor} = \left( \frac{P}{2 \pi} \right)^\frac{2}{3} \left( GM_\star \right)^\frac{1}{3} \label{eqn:rcor}
\end{equation} 
in orbit about the star, where $G$ is the gravitational constant. Values estimated for $R_\mathrm{cor}$ are listed in Table \ref{tab:starp} (column 13) and are typically 4--8 times the stellar radius (the largest ratio is 14.5 for EPIC 205151387).
With the luminosity of the star, we can estimate the effective temperature of material in a disk edge that is located at this radius from the star,
\begin{eqnarray}
T_\mathrm{cor}&=& 2^{-\frac{1}{2}}\, T_\mathrm{eff} \left(R_\star\over R_\mathrm{cor}\right)^{\frac{1}{2}} \label{eqn:tempa} \\
&=&0.34\, T_\mathrm{eff}\left(P\over 1\, \mathrm{day}\right)^{-\frac{1}{3}}\left(R_\star \over R_\odot\right)^\frac{1}{2}\left(M_\star\over M_\odot\right)^{-\frac{1}{6}},
\label{eqn:temp}
\end{eqnarray}
neglecting scattering, self-shielding or heat from accretion and assuming perfect blackbodies for the dust grains. These temperatures are listed in Table \ref{tab:starp}, column 12. 
Since the corotation temperature weakly depends on the mass, the error in the derived dust temperatures are similar to those for the stellar effective temperature error, i.e., $\pm200$ K. 
Blackbody approximation is good for dust grains $\geq1\,\mu$m but the temperature can be up to a factor of 2 larger for smaller grains \citep{2002Monnier}.
Inspection of this column shows that the disk temperatures at the corotation radius are below the dust sublimation temperature, $\sim 1400$ K \citep{2011Kobayashi}, indicating corotating gas is cold enough for dust to survive. Material corotating with the disk can cause dips in the light curve, through extinction from dust, if it is lifted above the disk midplane and can obscure the face of the star as viewed by the distant observer. 

Suppose the mechanism for forming dippers operates only at corotation. The material at the corotation radius must be below the dust sublimation temperature and this limits the rotation period. A star that is rotating quickly would have a corotation radius close to the star where the material is hot. If the stellar rotation period is short enough, dust could not exist at corotation. There is a minimum rotation period for a star of a given mass and luminosity allowing dust to survive at corotation. For a given stellar mass and effective temperature, a star's luminosity and radius depends on its age, so the temperature at the corotation radius (equation \ref{eqn:temp}) is a function of both stellar rotation period and age. We can invert Equation \ref{eqn:temp} to solve for the critical period at which the disk temperature at corotation is equal to the dust sublimation temperature as a function of stellar effective temperature. If the rotation period is above this critical period, dust can survive at corotation in the disk and if lifted above the midplane, the dust could obscure the star.

\begin{figure}
\includegraphics[width=3.3in, trim= 0 0 0 0 ]{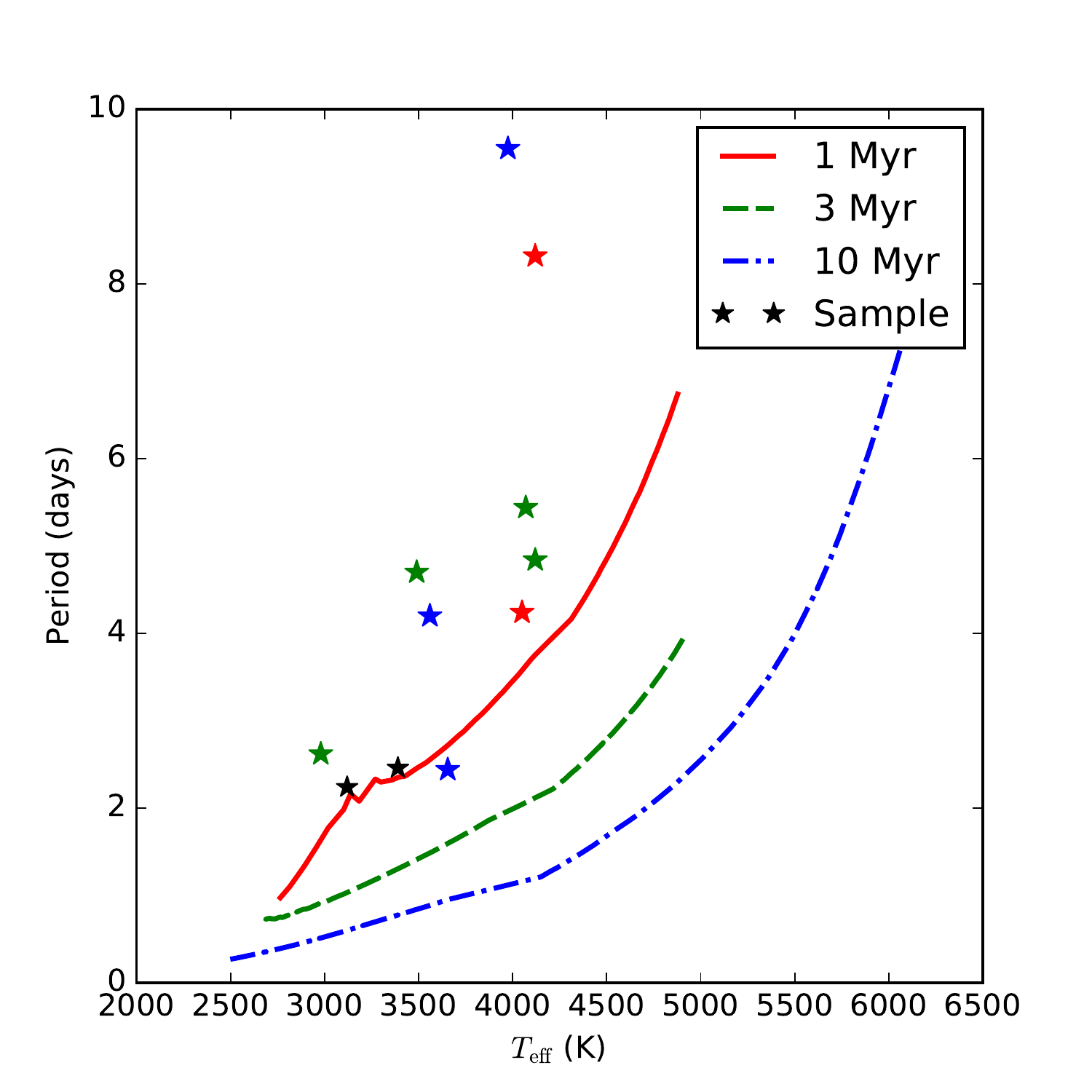}
\caption{The solid, dashed, and dash-dot lines mark the stellar rotation period at which the corotation radius of the disk is at 1400K (sublimating dust) for PARSEC  \citep{2012Bressan,2014Chen} pre-main-sequence isochrones of ages 1 Myr (red), 3 Myr (green), and 10 Myr (blue). The stars from our sample are marked by different coloured stars to indicate their age. Red points are for stars younger than 1Myr. Green are 1-3 Myr. Blue are 3-10 Myr and black is $>$10 Myr. Stars above the lines have corotation disk  temperatures cooler than 1400 K, allowing dust to survive.  We see that red points lie above the red curve, green points lie above the green curve and blue points lie above the blue curve.  All the stars shown on this plot are consistent with disk corotation below the dust sublimation temperature.  Dippers must lie above the curves shown here, so dippers are primarily restricted to cool, low-mass stars. 
Furthermore, higher-mass stars must have longer rotation periods than lower-mass stars to be dippers.
}
\label{fig:Temp}
\end{figure}

In Figure \ref{fig:Temp} we plot the critical period that allows dust to survive at corotation versus the stellar effective temperature, computed from stellar luminosities using three different PARSEC pre-main sequence tracks  \citep{2012Bressan,2014Chen}. The isochrones range from 0.09-1.5 $M_\odot$ and we show minimum periods computed using isochrone ages 1 Myr, 3 Myr and 10 Myr as solid red, dashed green, and dash-dot blue lines. The minimum period is computed by solving equation \ref{eqn:temp} for $P$ and using a temperature limit $T_{lim} = 1400$K for dust sublimation. Above the red curve for 1 million year old stars, the green curve for 3 million year old stars, and the blue curve for the 10 million year old stars, we can see which corotation periods allow dust to survive. In Figure \ref{fig:Temp} stars from Table \ref{tab:starp} are plotted as points with colours denoting their ages. We see that red points (stars at or below 1 Myr) are above the red curve, and similarly green points (stars 1-3 Myr old) are above the green curve and blue points (3-10 Myr year old stars) are above the blue curve, consistent with the corotation temperatures being cool enough for dust survival. The black points are for stars older than 10 Myr and since the curves are lower with increasing age, these points are also consistent with corotation temperatures cool enough for dust survival.

For the lower-mass, cool stars, the minimum period allowing dust to exist at corotation is around a couple days. Figure \ref{fig:Temp} shows that the minimum period allowing dust to survive at corotation quickly increases with stellar effective temperatures above $\sim$4000 K. The hotter the star, the higher the curves. This means that hot stars must have very long rotation periods for the corotating disk to be cold enough to host dust. Only cooler, older stars allow quickly-rotating stars to host dust at corotation in the disk. Dippers must lie above the curves shown here, so dippers are primarily restricted to cool, low-mass young stars. 

\cite{2016A} found a significant correlation (coefficient of 0.77) between the rotation period and the effective temperature of the star for their sample of ten dippers. We can explain this correlation if the dipper phenomenon is preferentially associated with dusty material corotating with the star.  Furthermore, Figure \ref{fig:Temp} shows that only low-mass stars allow corotating dust to survive.  The hypothesis that dippers are associated with corotating material explains why dippers are preferentially low-mass stars.

Dust near its sublimation temperature emits strongly in the near infrared. All 10 stars discussed by \citet{2016A} display mid-infrared excesses.  \cite{2016A} found a moderate correlation (coefficient $\sim$0.5) between the largest dip depths for each star and the excess in the 4.6~$\mu$m band compared to K-band (2.15 $\mu$m). So the stars with the largest dips likely have dusty disks that extend into corotation.
However, some stars show little or no excess at 3.4~$\mu$m or 4.6~$\mu$m yet have flat or rising excesses at longer infrared wavelengths, indicating they host (pre-)transition disks. Late-K and M dwarfs are bright at wavelengths shorter than $5\,\mu$m which makes determining the excess (from disk emission) in these bands difficult.  A hot inner dusty disk is probably not ruled out by the spectral energy distributions.
%Hence most but not all of the 10 stars discussed by \citet{2016A} The spectral energy distributions and associated models by \citet{2016A} were consistent with that suggest that their inner disks extend into the corotation region. However, their SED modelling was simplistic, ignoring gaps and suffered from some degeneracy. An inner disk followed by an annular gap and then an outer disk was not accounted for in the modelling. An inner disk may be masked by the star at wavelengths $<5\,\mu$m and the gap would cause a dip in the SED which could push the modelled inner disk radius beyond the corotation radius.

We check that the prototype dipper, AA~Tau, has a similar corotation disk temperature by comparing it to the stars in Table \ref{tab:starp}. It has a spectral type K5, effective temperature 4060 K, mass $0.8\, M_\odot $, radius $1.8\, R_\odot$ \citep{gudel07},  accretion rate a few times $10^{-9}\,  M_\odot {\rm yr}^{-1}$ and rotation period  8.2 days \citep{bouvier03}, making it similar to the hotter stars listed in Table \ref{tab:starp}.  AA Tau has a similar mass and radius as EPIC~203937317 but a longer period implying that its corotation temperature would be cooler than that of EPIC~203937317. We have estimated the corotation temperature for AA Tau to be $T_\mathrm{cor} \sim 1150$ K.

\subsection{Light curve slopes} \label{sec:slopes}

\begin{figure*}
\includegraphics[width=6.6in, trim= 0 0 0 0 ]{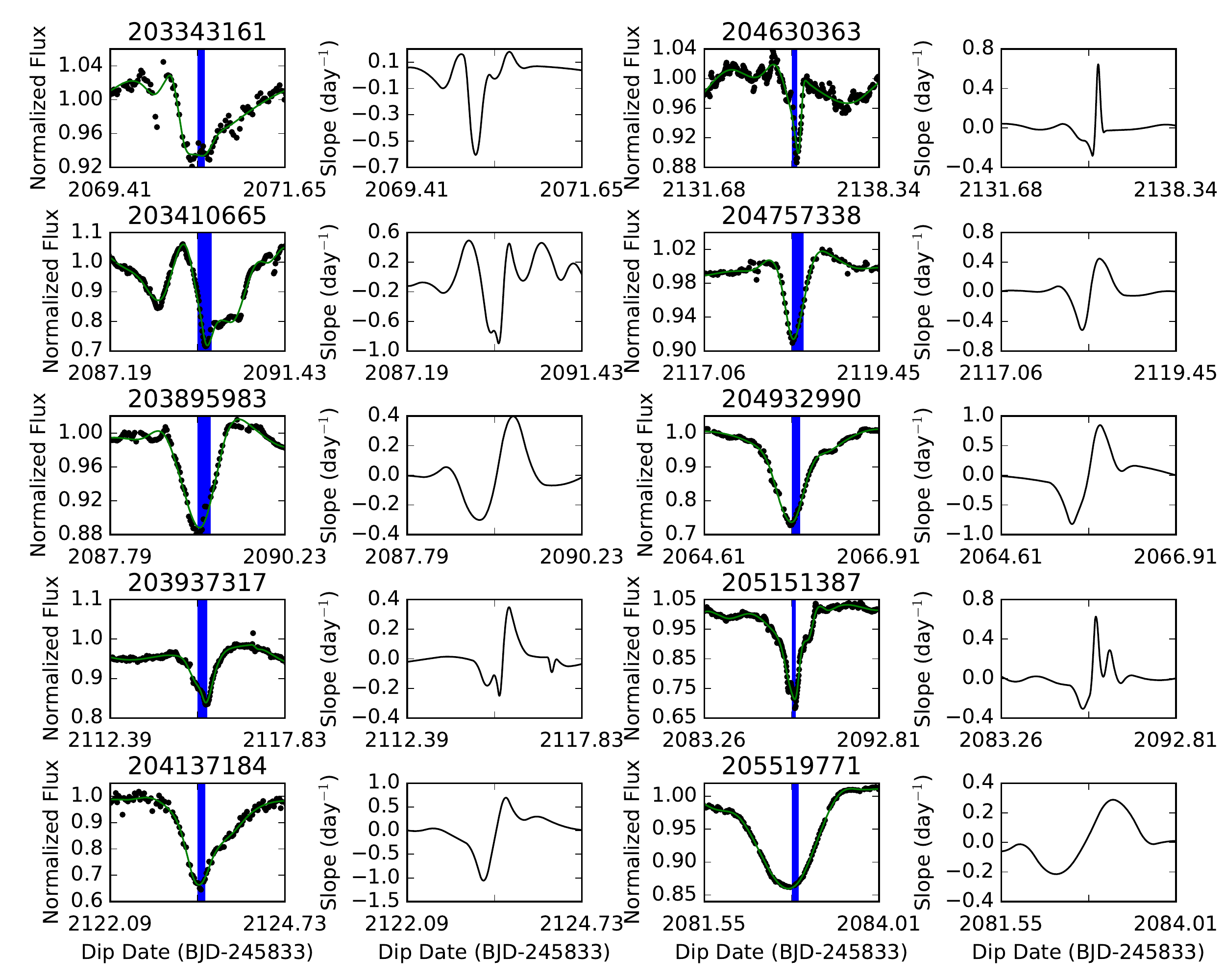}
\caption{Each pair of plots shows a single rotation period for a single star with light curves on the left and slope (change in flux per day) as a function of time on the right.  The blue or red bars have widths equal to the crossing timescale ($t_\mathrm{cross}$, equation \ref{eqn:tcross}) for a small object crossing in front of the star at the corotation radius. The blue bars are shown for stars with estimated stellar and corotation radii, otherwise red bars are shown assuming $R_\mathrm{cor}/R_\star=5$.  When the dips are wider than the crossing timescale, the disk structure or accretion stream passing in front of the star must be wider than the star's diameter.  Maximum slope amplitudes are measured from the slope curves and used to test the hypothesis that they are consistent with material orbiting at the corotation radius.
}
\label{fig:dip1}
\end{figure*}
\setcounter{figure}{2}

\begin{figure*}
\includegraphics[width=6.6in, trim= 0 0 0 0 ]{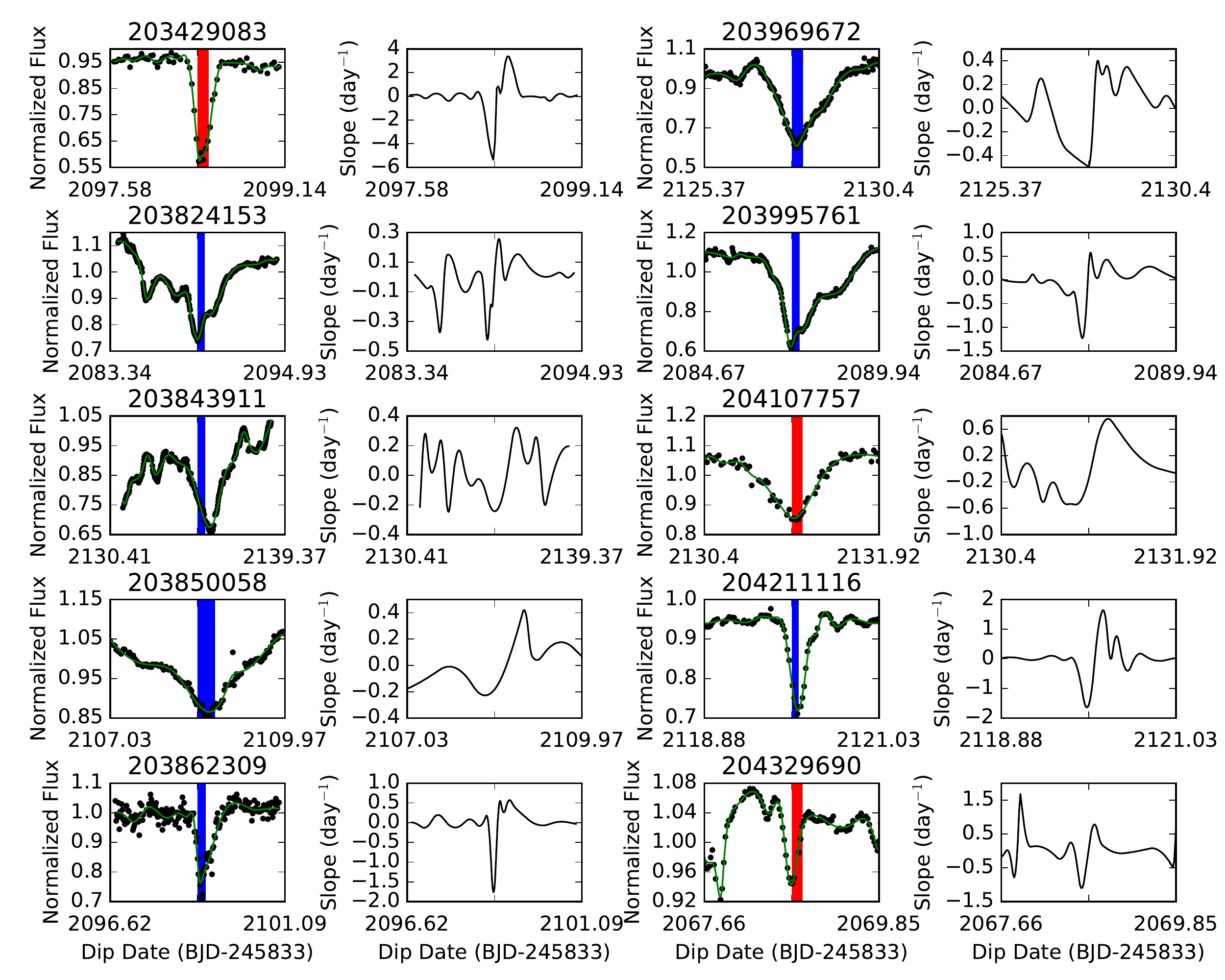}
\caption{continued.
}
%\label{fig:dip2}
\end{figure*}
\setcounter{figure}{2}

\begin{figure*}
\includegraphics[width=3.3in, trim= 0 0 0 0 ]{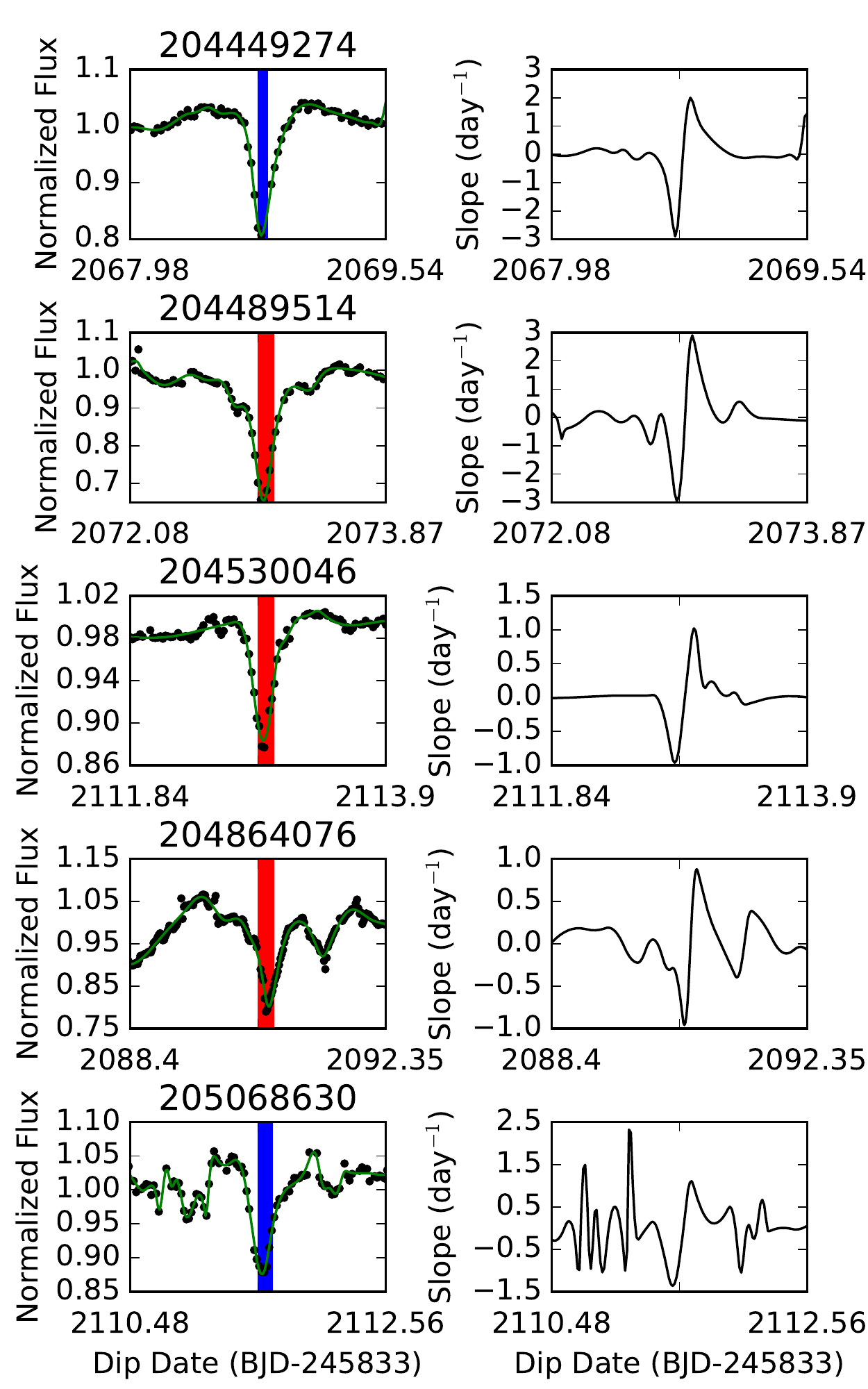}
\caption{continued.
}
%\label{fig:dip3}
\end{figure*}

We can test our hypothesis that dippers are associated with corotating material with the light curve slopes. An object that occults a star causes a drop in the flux at a rate that depends on the velocity of that object. If the material obscuring the star is corotating with the star then we can compute a limit on the light curve slope (change in light curve flux per day) associated with the rotation speed at the corotation radius $R_\mathrm{cor}$.
 
A large optically-thick rectangular block rotating at Keplerian circular velocity, 
\begin{equation}
V_\mathrm{cor}  =\sqrt{GM_\star/R_\mathrm{cor}}, \label{eqn:Vcor}
\end{equation} 
at corotation, that completely occults the star, causes a maximum light curve slope with absolute value $s_\mathrm{max} \approx V_\mathrm{cor}/(2 R_\star)$ where $2 R_\star$ is the diameter of the star and stellar flux is normalised so that the light curve has a value of 1 when the star is not occulted.

A very small object (compared to $R_\star$)  in orbit about the star at radius $R_\mathrm{cor}$ occults the star a fraction of the rotation period $2 R_\star /(2 \pi R_\mathrm{cor} )$ where we have divided the diameter of the star $2 R_\star $ by the orbit circumference $2 \pi R_\mathrm{cor}$ and assumed that the object passes over the star's equator in the observer's view. The time that it takes a small object to pass across the diameter of the star we call the crossing time,
\begin{equation}
t_\mathrm{cross} \equiv P \frac{R_\star}{\pi R_\mathrm{cor}}. \label{eqn:tcross}
\end{equation}
When a large optically-thick block is narrow compared to the star radius, it does not ever completely cover the star. The maximum slope in the light curve due to the block at corotation is
\begin{equation}
s_\mathrm{max} \approx \frac{\pi\delta}{P} \frac{R_\mathrm{cor}}{R_\star} = \frac{\delta}{t_\mathrm{cross}} \label{eqn:smax}
\end{equation}
where $\delta$ is the depth of the dip normalised to the star's total flux and approximately corrects for the occulter opacity (if not optically thick) and the fraction of the star surface covered during the light curve minimum.
%We compute the crossing times from equation \ref{eqn:tcross} using the rotation periods and stellar radii listed in Table \ref{tab:starp}.
A comparison of the predicted maximum slope (from equation \ref{eqn:smax}) to the maximum slope measured in the light curves will test the hypothesis that dippers are associated with corotating material.

The slope of a light curve can be computed from the difference in fluxes at adjacent times.  However, noise in the measured light curve introduces errors into this computation.  To reduce the error caused by noise and short term stellar variability we fit a cubic B-spline (or basis spline) to each light curve using the python routine \texttt{scipy.interpolate.splrep} \citep{1993Dierckx}. A smoothing condition was adjusted for each light curve to minimise short term variability while maintaining the depth of the dips.  For a single rotation period for each star, we show the normalised light curve (as points) along with the spline fits (shown as green lines) in Figure \ref{fig:dip1}. To compute the light curve slopes we used the accompanying routine \texttt{scipy.interpolate.splev} giving the derivative of the B-spline fit. The slopes in units of normalised flux per day for each star are plotted to the right of the light curves in Figures \ref{fig:dip1}. We only show a single rotation period for each star, choosing a period exhibiting one of the deepest dips.  Blue and red bars are shown in Figure \ref{fig:dip1} with width equal to the crossing timescale computed using equation \ref{eqn:tcross}. The bar is blue if the crossing timescale is computed using an estimate of the stellar radius when available (in Table \ref{tab:starp}) and shown in red assuming $R_\mathrm{cor}/R_\star=5$ otherwise.

  Many of the light curves have dips that are much wider than the blue or red bar. These we interpret as disk structures from wide accretion streams or a disk warp, occulting structures that are wider than the diameter of the star. A few stars show dips that have widths about as narrow as the crossing time (e.g., EPIC~204630363). These have occulting streams that are narrower than the diameter of the star.  
 The crossing timescale also shows how the diameter of the star affects the light curve.  Fine features in the disk cannot easily be seen in the light curve because the viewer detects light from the entire surface of the star and so measures a sum that is sensitive to all structures that occult the face of the star at a single time. The vertical bars give us an estimate of how the diameter of the star effectively smooths over the occulting disk or stream structure. 

\begin{figure}
\includegraphics[width=3.3in, trim= 0 0 0 0 ]{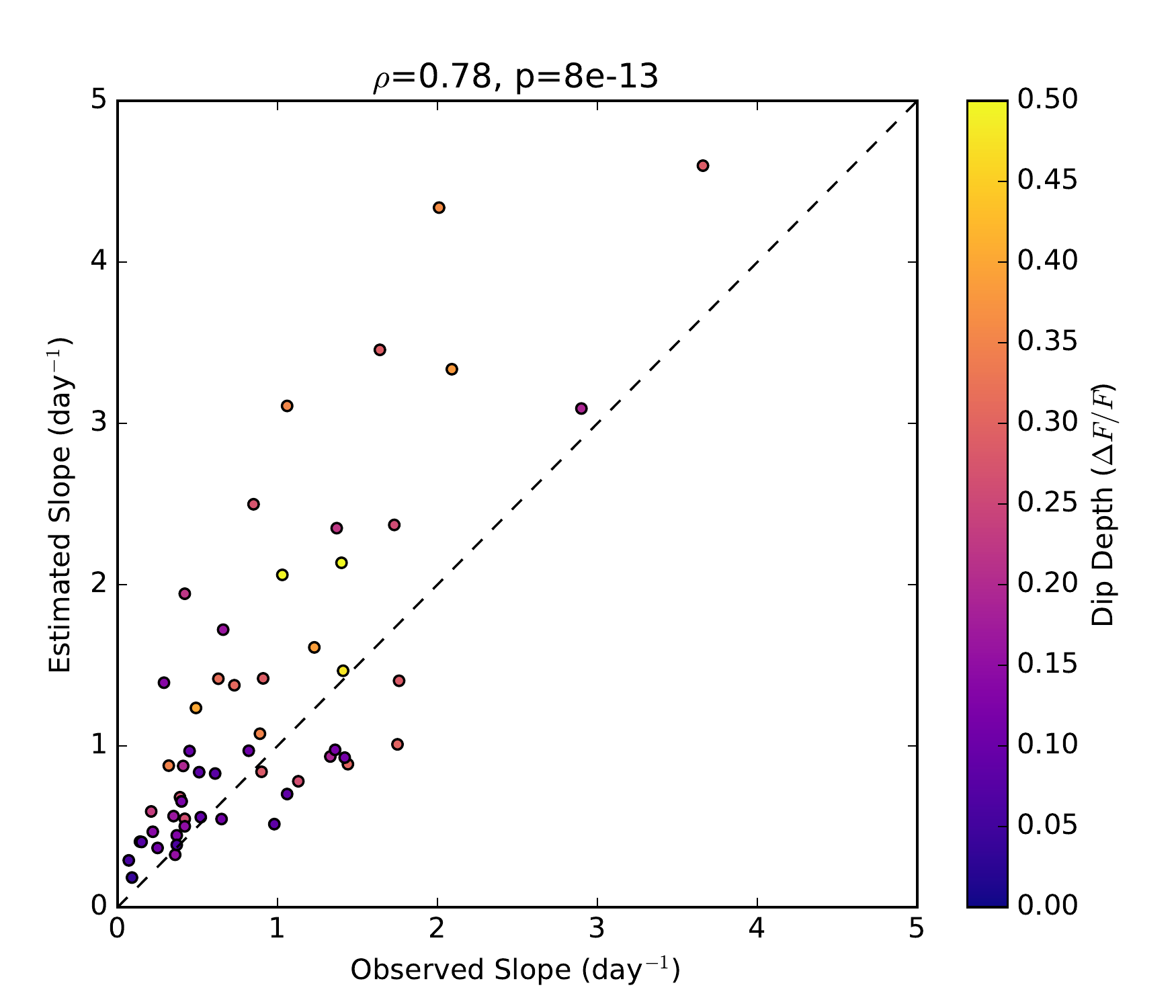}
\caption{Comparison of the observed maximum light curve slopes versus the computed values. The dots are coloured to indicate dip depth.  The  dashed line shows where the observed and computed slopes are equivalent. The Spearman rank correlation coefficient ($\rho$) and the p-value for the points are listed above the plot. The correlation is strong, supporting our hypothesis that material causing the dips is corotating with the star.
}
\label{fig:slope}
\end{figure}

For each star, we measure the maximum absolute value of the slopes from the three deepest dips. These we refer to as measured values of $s_{max}$. We then compute the maximum slope using equation \ref{eqn:smax}, the star and corotation radii in Table \ref{tab:starp} and the depth of the dip. This we refer to as the computed value of $s_{max}$. On Figure \ref{fig:slope} we plot the computed $s_{max}$ (on the y-axis) against the measured values (on the x-axis). The computed and measured slopes are well correlated with a Spearman rank correlation coefficient of 0.78 ($p$-value=$10^{-12}$). The strong correlation supports our hypothesis that the dust occulting the star is near the corotation radius. 

Most of the observed slopes are within a factor of two of the computed values.  Many of the dips are wider than a crossing timescale (see Figure \ref{fig:slope}) and this implies that the occulting object is much wider than the stellar diameter. For many of the stars, the computed maximum slope exceeds the maximum measured one, as would be expected if the occulting material is wider than a stellar radius and badly approximated by a narrow opaque block. This is consistent with the fact that most dips' widths exceed their associated crossing timescales. 

%How uncertain are the computed maximum slopes? 
Several dips have measured maximum slopes larger than the computed one. However, only points significantly below the line would present a strong contradiction to our hypothesis due to the errors in computed slopes. Stellar radii, $R_\star \propto L_\mathrm{Bol}^\frac{1}{2} T_\mathrm{eff}^{-2}$, have uncertainty primarily dependent on errors in the effective temperature. An error of $\Delta T \sim 200 K$ gives  only a 10\% error in $R_\star$.   There are large systematic errors in the stellar masses as we used stellar isochrones to estimate them \citep{2004Hillenbrand}, but the corotation radius, $R_\mathrm{cor}$, is only weakly dependent on stellar mass (to the 1/3 power;  see equation \ref{eqn:rcor}). Using equations \ref{eqn:tcross} and \ref{eqn:smax} and assuming the uncertainty arises from a 200 K error in effective temperature, we estimate an error of $\sim$10\% in the computed maximum slope.  Therefore, points slightly below the line in Figure \ref{fig:slope} can be attributed to errors in the stellar effective temperature. % A small error in the measured slopes occurs since we did not remove the star spot fluctuations. 
A higher slope in the light curve would be seen when a stream rotates in front of the star surface at the same time as a star spot rotates below the horizon. 

\subsection{The Magnetospheric Truncation Radius}\label{sec:mag}

An accreting circumstellar disk can be truncated by the magnetosphere associated with the stellar magnetic field \citep{camenzind90,konigl91}. We consider the possibility that the corotation radius (for the dippers) is near the magnetospheric truncation radius. 
%Ten of the stars have X-ray counterparts found in the HEASARC catalogues that confirm the stars' high magnetic activity. 
The radius of magnetospheric truncation can be estimated by balancing magnetic pressure from the stellar magnetic dipole component against ram pressure in the accreting disk.  Assuming spherical accretion, the truncation radius is
\begin{eqnarray}
{R_T\over R_\star}&=&7.1 B_3^\frac{4}{7} \dot{M}_{-8}^{-\frac{2}{7}}
 \left(M_\star\over 0.5\, M_\odot\right)^{-\frac{1}{7}}\left(R_\star \over 2\,R_\odot\right)^\frac{5}{7}
\label{eqn:trunc}
\end{eqnarray}
(equation 2 by \citealt{2007Bouvier})
where $B_3$ is the dipole magnetic field strength at the stellar photosphere in units of kG and $\dot{M}_{-8}$ is the accretion rate in the disk in units of $10^{-8} M_\odot\,yr^{-1}$. Interactions with a circumstellar disk are, in most cases, dominated by the large scale dipole field \citep{2008Gregory,2012Adams} as the quadrupolar and higher components decay more rapidly with distance from the star.  A similar relation is derived by matching the azimuthal components of the stress tensor (e.g., see section 2.2 in review by \citealt{romanova15} and \citealt{konigl91}). 

\begin{figure}
\includegraphics[width=3.3in, trim= 0 0 0 0 ]{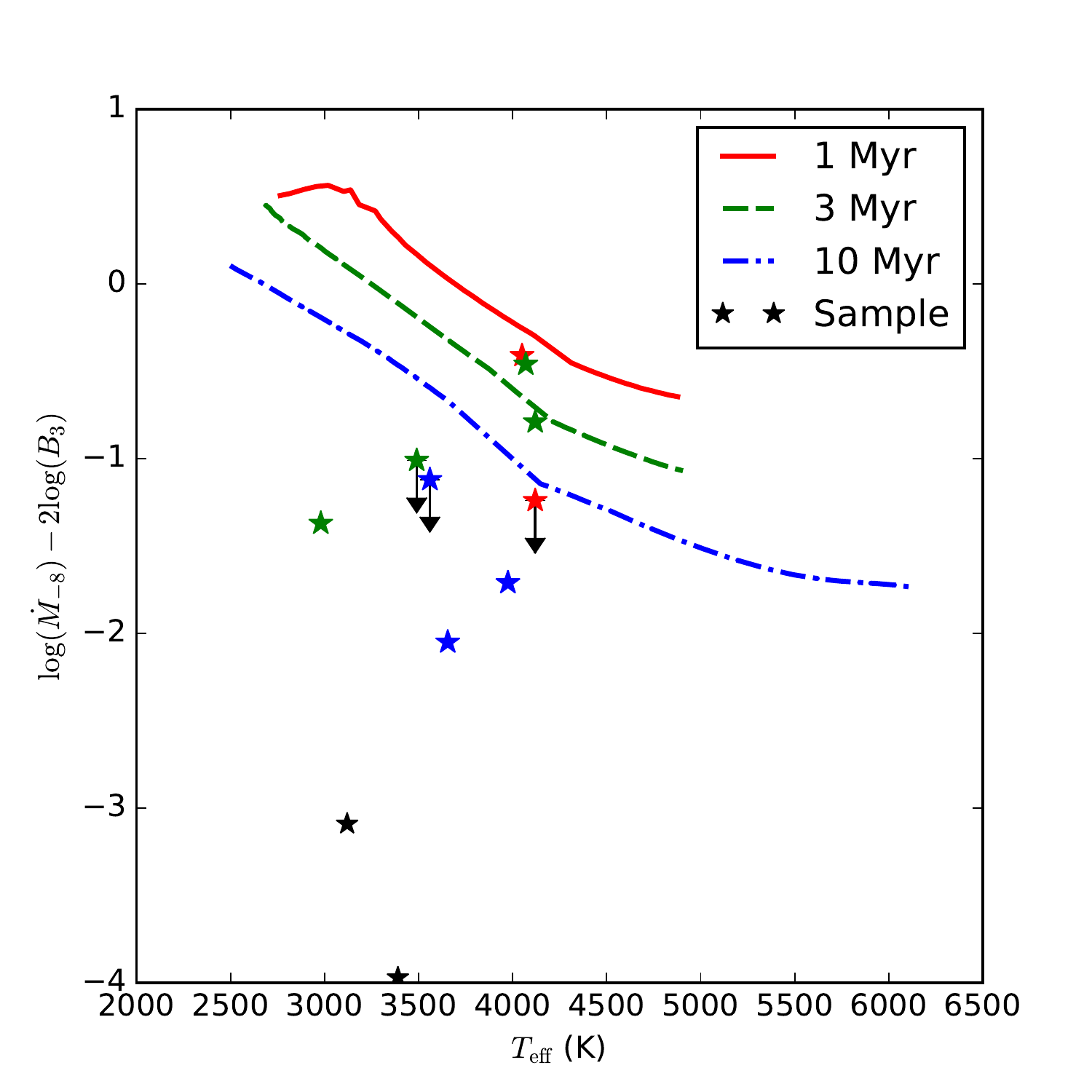}
\caption{Upper limits on accretion rate divided by the square of the surface dipole magnetic field strength for a magnetospheric truncation radius that is 1000 K.  The lines show limits using isochrones for 1 Myr, 3 Myr and 10 Myr, in solid red, dashed green and dash-dot blue, respectively. Stars with accretion rates listed in Table \ref{tab:acc} are plotted as coloured points on the plot using a magnetic field of 1 kG to calculate their $y$ values. The star ages are coloured in the same way as in Figure \ref{fig:Temp}. The regions above the curves have magnetospheric truncation radii at higher temperature than 1000 K, whereas those below the curves have magnetospheric truncation radii below 1000 K. As the dippers in Upper Sco and $\rho$-Oph have corotation temperatures of $\sim$1000 K, we expect them to lie near the curves. Regions below the curves could exhibit dipper phenomena as their disk truncation radii would be cool enough to allow dust to survive. Regions above the curves should be devoid of dippers as truncation radii are so hot dust should have sublimated in the accretion streams. If surface dipole fields are a factor of a few below 1kG then the points would lie nearer the curves.
}
\label{fig:acc}
\end{figure}

Above (see sections \ref{sec:spec} -\ref{sec:slopes}) we have associated dips in the light curve with a radius in the disk that is corotating with the star and found that this material has temperature approximately 1000K. The fact that the temperature at the corotation radius must be below the dust sublimation temperature puts a limit on the rotation period of the star.  By associating the magnetospheric truncation radius with a specific temperature we can derive a similar limit but on the accretion rate and stellar magnetic field strength. Using an assumed disk edge temperature, we compute its radial location with equation \ref{eqn:temp}. This we insert into equation \ref{eqn:trunc} to solve for the accretion rate times a factor of the surface dipole field strength $B^{-2}$. The result is 
\begin{eqnarray}
\dot{M}_{-8}B^{-2}_3&\sim &950\left(R_T\over R_\star\right)^{-\frac{7}{2}}\left(M_\star\over 0.5\, M_\odot\right)^{-\frac{1}{2}}\left(R_\star \over 2\,R_\odot\right)^\frac{5}{2} \nonumber \\
&\sim &11000\left(T_\mathrm{disk}\over T_\mathrm{eff}\right)^{7} \left(M_\star\over 0.5\, M_\odot\right)^{-\frac{1}{2}}\left(R_\star\over 2\,R_\odot\right)^\frac{5}{2}.  \label{eqn:MB}
\end{eqnarray}
 In the second line, we have used equation \ref{eqn:tempa} to replace $R_T$ with $T_\mathrm{disk}$, but for the disk temperature at the truncation radius. We plot the limit on $\dot{M}_{-8}B^{-2}_3$ using a disk temperature of $T_\mathrm{disk} = 1000$ K in Figure \ref{fig:acc} and using the same isochrones as for Figure \ref{fig:Temp}. 

If the stellar magnetic field is weak or the accretion rate is high, so that $\dot{M}_{-8}B^{-2}_3$ is above the curves on Figure \ref{fig:acc}, then the disk can extend closer to the star and reach higher temperatures, above the dust sublimation temperature. Accretion streams or a warp associated with such an inner disk should not contain dust. We do not expect dippers associated with streams to lie above these curves because if dust is sublimated in the inner disk, we would not see it in extinction against the star.

Our dippers should lie near the curves in Figure \ref{fig:acc} because we have inferred that their disks extend to approximately 1000K. Truncated disk edges could be even cooler than 1000K if outside the corotation radius, so stars below the line might still exhibit dipper phenomena. The lines in Figure \ref{fig:acc} give upper limits to the dipper phenomena and imply that dippers should primarily be stars with low accretion rates. This is consistent with the late stage of disk evolution inferred by the spectral energy distributions of dippers in the sample and the low accretion rates inferred from the weak equivalent widths in H$\alpha$ and Pa$\gamma$ \citep{2016A}.

Stars above the lines in Figure \ref{fig:acc} may show dips in their light curves but the obscuring dust cannot be near the magnetospheric truncation radius and also likely not near the corotation radius like the stars in this sample. An example of a dipper that would lie above the line is HD~142666, a UX~Orionis-type variable star in Upper Sco \citep{1998Meeus}. The light curve of HD~142666 displays dips similar to the extinction dominated variability in this sample. However, the dips occur on timescales of days which is much longer than its rotation period of $\sim0.2$ days \citep{2014Vural}. HD~142666 also has a high accretion rate \citep{2011Mendigutia} so the magnetospheric truncation radius is small and within the dust sublimation radius. This is unlike the stars in our sample which display dips associated with material near the star's corotation radius and have low accretion rates. The mechanism creating the dips in flux observed in UX~Orionis-type stars is different from the one causing the dipper phenomena in our sample.

 Many of the stars in the sample are classified by \citet{2016A} as weak T~Tauri stars (WTTS) based upon their H$\alpha$ emission. WTTS are widely assumed to be longer accreting and have evolved debris disks so they should not form streams and warps at the disk edge. However, \citet{2016A} found weak accretion signatures for most of the WTTS in their sample indicating that the dippers are near the CTTS-WTTS transition. We expect the dippers in our sample are at least weakly accreting so that accretion streams or warps are forming.
 
There are six stars in the sample with measured accretion rates or upper limits \citep{2006Natta} which we list in Table \ref{tab:acc}. \citet{2016A} did not estimate accretion rates because they are highly uncertain. For example, EPIC~203850058 has two other reported accretion rates in the literature other than the one listed in Table \ref{tab:acc}: $\log \dot{M} = -9.0$ \citep{2004Natta} and -9.98 \citep{2015Manara} with $\dot M$ in units of $M_\odot $~yr$^{-1}$.  \cite{2006Natta} uses an empirical relation to estimate the accretion luminosity from Pa$\beta$ equivalent widths, $\log L_\mathrm{acc}=1.36(\pm0.2)\times\log L({\rm Pa}\beta)+4.00(\pm 0.2)$, from which an accretion rate is computed ($\dot{M}=L_\mathrm{acc}R_\star/(GM_\star)$). Since the equivalent widths of Pa$\gamma$ are within a factor of a few of Pa$\beta$ equivalent widths \citep{2014Smith}, we can use the Pa$\gamma$ measurements from \cite{2016A} and the same empirical relation used by \cite{2006Natta} for Pa$\beta$ to crudely estimate the accretion rates of five more stars. These accretion rates are also listed in Table \ref{tab:acc}. 
 
 We place the stars in Table \ref{tab:acc} with estimates of accretion rates onto Figure \ref{fig:acc} with colours to indicate age the same way as in Figure \ref{fig:Temp}. When measured accretion rate is an upper limit, we have plotted a point and a black arrow in Figure \ref{fig:acc}. The $y$ values for each point are computed using the estimated accretion rates and a dipole surface magnetic field strength of 1kG,  and the $x$ values are the effective stellar temperatures from Table \ref{tab:starp}.  Figure \ref{fig:acc} confirms that all but one star lies below the limiting curves. One star (EPIC~203937317, a green point) lies above the accompanying green line, but its corotation disk temperature, estimated at 1150 K, is slightly higher than 1000 K and had we used a higher temperature to estimate the limit, the star would have been under the limit. The limiting curves for  $\dot{M}_{-8}B^{-2}_3$ depend on the temperature to the 7th power (equation \ref{eqn:MB}) and so are sensitive to the disk temperature used to compute the limit. 

%The corotation radius cannot be far outside the magnetospheric truncation radius because then the inner disk can accrete until it extends to smaller radii where dust can sublimate and we would no longer see a dipper. Corotation radii cannot be significantly smaller than the magnetospheric truncation radius since the disk would have been truncated and we would not see dipper phenomena associated with corotation. Furthermore, if the magnetospheric truncation radii were larger than the corotation radius then the system would be in what is called the ``propeller'' regime where outflows are possible but accretion is not (see section 2.2 and figure 4 by \citealt{romanova15}).

We expect the stars to lie near the limiting curves on Figure \ref{fig:acc}, but many of the stars are well below these curves. Either the estimated accretion rates are too low, the dipole component of the stellar surface magnetic fields are lower than a kG, or equation \ref{eqn:trunc} should be modified. However, without magnetic field measurements and with the large uncertainties in accretion rate, we cannot discern which is the cause of the discrepancy.

The accretion rates listed in Table \ref{tab:acc} are not unexpected compared to similar low mass stars in the same region (e.g., \citealt{2006Natta}). Using ram pressure in a thin disk rather than spherical accretion, equation \ref{eqn:trunc} gains an additional factor of $\alpha^{-2}(h/R_T)^{6/7}$ \citep{2001Blackman} where $\alpha$ is a viscosity parameter and $h$ is the scale height. The additional factor can reduce the truncation radius, lowering the lines shown in Figure \ref{fig:acc}, possibly bringing the points for the stars into closer agreement with their expected locations near the lines, though an estimate for $\alpha$ would be required to do this.

Because the small-scale field components decay faster with height above the stellar surface, the dipole component is thought to be most important for magnetospheric disk truncation \citep{konigl91}. Derived from the technique of Zeeman-Doppler imaging, magnetic surface maps have now been published for a number of accreting T~Tauri stars and these have been used to estimate the strength of the dipole component of the surface magnetic field (e.g., \citealt{gregory12}). \cite{gregory12} found empirical regions on the HR diagram to predict global properties of a pre-main-sequence star's magnetic field; see their Figure 4. Most of the stars in our sample are in regions on the HR diagram where the magnetic field is axisymmetric and the dipole field is strong. However, several, including the two very low points in Figure \ref{fig:acc}, are in what \cite{gregory12} labeled region 4, which is populated by stars with a mixture of magnetic field topologies. These stars can have a strong dipole field or a complex, non-axisymmetric field with a weak dipole component that is significantly less than the 1 kG estimate used in our calculations. Decreasing the surface dipole magnetic field strength by a factor of a few moves the stars in Figure \ref{fig:acc} upwards by about 1, bringing many of the plotted stars in proximity to the curves. 

Ten of the Table 1 stars have X-ray counterparts in the ROSAT, XMM, and/or Chandra catalogs available on HEASARC\footnotemark. Most of the detected stars are ROSAT all-sky survey detections and, as expected (since the RASS sensitivity was insufficient to detect young M stars much beyond $\sim$50 pc; e.g. \citealt{2016Kastner}), the majority (8) of these ROSAT detections are K stars. The only Table 1 K star that went undetected in the RASS is EPIC~204630363. There is also a handful of XMM and Chandra serendipitous detections of Table 1 stars in the HEARARC database. The fact that the majority of the Table 1 K stars are detected in the RASS, and that a few of the M stars are serendipitous Chandra and XMM sources, supports our working assumption that these are young K/M stars that are highly magnetically active. 
\footnotetext{https://heasarc.gsfc.nasa.gov/cgi-bin/W3Browse/w3browse.pl}

Because we lack accurate measurements of accretion rates and any measurements for the stellar dipole magnetic field strengths we cannot definitively say whether the corotation radii are equivalent to magnetospheric truncation radii. However, the estimated magnetospheric truncation radii are consistent with being \textit{near} the corotation radii, within our large level of uncertainty. We do not expect dippers from accretion streams or warps to be observed if the magnetospheric truncation radius were significantly smaller than corotation since the dust would not survive. If the magnetospheric truncation radii were larger than the corotation radius then the star is in the ``propeller'' regime (see Figure 4 by \citealt{romanova15}). One possibility is that the magnetospheric truncation radius is near but somewhat smaller than corotation, $R_\mathrm{cor} /R_T \sim 1.2 - 1.3$, corresponding to the equilibrium state seen in simulations \citep{2005Long}. More observations to measure magnetic dipole field strength and accretion rate are needed to estimate the magnetospheric truncation radius to corotation radius ratio.

\begin{table}
\begin{center}
\vbox to 70mm{\vfil
\caption{\large Observed Accretion Rates}
\begin{tabular}{@{}lcc}
EPIC 		& $\log_{10}(\dot{M})$ & Ref. \\
	& $\dot{M}\,yr^{-1}$	& \\
\hline
203343161	& -11.09		& 2 \\
203410665	& -8.41		& 2 \\
203895983	& -10.05		& 2 \\
203937317	& -8.46		& 1 \\
205151387	& -9.71		& 2 \\
205519771	& -11.97		& 2 \\
\hline
203843911 	& $<$-9.24 	& 1 \\
203850058	& -9.37 		& 1 \\
203862309	& $<$-9.12	& 1 \\
203969672	& $<$-9.01	& 1 \\
203995761	& -8.79 		& 1 \\
\hline
\end{tabular}
{\\References: (1) \cite{2006Natta}
(2) Estimated using the EW(Pa$_\gamma$) measured by \cite{2016A}.  Accretion rates ($\dot M$) are in solar masses per year. \\
\label{tab:acc}}
\vfil}
\end{center}
\end{table}

\subsection{The Evolution of Corotation and Magnetospheric Truncation Radii}  \label{sec:trunc}

Star-disk magnetic interactions and magnetised winds and outflows remove angular momentum from actively-accreting young stars \citep{matt05,davies14}. Stellar field lines that couple to the disk outside the corotation radius slow the rotation of the star down, while field lines that couple to the disk inside corotation will speed the rotation up \citep{ghosh77,2007Bouvier}. Therefore, the value of the truncation radius ($R_T$) relative to the corotation radius ($R_\mathrm{cor}$) determines whether the star spins up or down. When the corotation radius is larger than the truncation radius, the star spins up.

The value of $R_T$ relative to $R_\mathrm{cor}$ also determines whether material can accrete onto the star. Within the truncation radius, disk material is locked to the stellar field lines and moves at the same angular velocity as the star. If $R_T < R_\mathrm{cor}$, then the disk material will lose angular momentum to the magnetic field and accrete onto the star. This results in the star spinning up. For $R_T >  R_\mathrm{cor}$, material locked to the stellar magnetic field will be moving faster than the local Keplerian velocity in the ``propeller'' regime \citep[e.g.,][]{1999Lovelace} and so will be pushed away from the star gaining angular momentum from the star and slowing its rotation. \cite{2005Long} shows that a star can reach an equilibrium state where it neither spins up nor down when the star rotates slightly slower than the inner disk.

 Consider a disk initially with truncation radius corotating with the star, $R_T \sim R_\mathrm{cor}$. Suppose that the disk accretion rate can vary on a short  timescale compared to the time it takes to change the stellar rotation rate or magnetic field strength. The magnetospheric truncation radius varies with the accretion rate. If the accretion rate drops, the magnetic truncation radius must move outwards. At this point $R_T > R_\mathrm{cor}$ and material will be flung away from the star. However if this state is on average maintained over longer timescales, the star will spin down and eventually the corotation radius would move outwards allowing $R_\mathrm{cor}$ to match $R_T$. If the accretion rate increases, then the magnetic truncation radius must move inwards so that $R_T < R_\mathrm{cor}$.  In this state accretion onto the star is allowed, but then then star can eventually spin up if this state is maintained. Again, the corotation will approach the disk truncation radius. 
If the propeller regime or accretion regime is maintained for a timescale comparable to the timescale for stellar rotation to vary then the corotation radius and disk truncation radius may approximately coincide.
 Measurement of accretion rates and magnetic field strength would test the hypothesis that the truncation radii are near corotation in dippers and thus the scenarios for coupled disk and star rotational evolution.

Because of this evolution of the stellar spin, we expect the truncation radius and the corotation radius of the stars in our sample to be similar. The dippers phenomenon can occur when the truncation radius is smaller but still near the corotation radius so the star is accreting. If the dust obscuring the star is associated with warps and streams along the inner edge of the disk, then the truncation radius cannot be much smaller than the corotation radius otherwise we would have detected multiple periods in some of the spectrograms. If the dust is not associated with the disk edge, then the inner edge can be within the dust sublimation radius but we would not expect all the periodic dippers with rotation signatures in our sample to be corotating with the star. Spectroscopic measurements of the stellar rotation period are needed to confirm the spectrograms.

\subsection{Lifting Material Out of the Midplane}\label{sec:lift}

Magnetic surface maps of CTTS have shown that misalignment between the star's magnetic field and the star's rotation axis is common \citep{gregory12}. Material in the disk midplane when it reaches the magnetospheric truncation radius can no longer drift inwards and so will pile up there. Once the pressure is high enough, disk material is pushed upwards and downwards along the field lines and out of the disk plane forming accretion funnels \citep{romanova03}.  Depending on tilt axis and field strength, the inner disk can form a warp  (as in AA~Tau, \citealt{bouvier99}) or accretion streams (see \citealt{romanova15} for review and \citealt{stauffer15} for observational evidence). Warps and streams can occur together if the star and disk edge are nearly corotating\citep{romanova13}. In both cases, the magnetosphere of the star is responsible for lifting material out of the midplane. 
Previous observational studies have shown results that agree with this mechanism including AA~Tau \citep{bouvier99,bouvier03} and NGC~2264 \citep{2010Alencar, 2015McGinnis, stauffer15}.

Above we began by showing that, for most dippers, the dipper period is similar to the stellar rotation period and so associated with a corotation region in the disk. Using estimates for magnetic fields and accretion rates, we found that the magnetospheric truncation radii could be close to the corotation radius 
but the truncation radii are highly uncertain. The association is important because if magnetospheric truncation radii are equal to the corotation radii, the magnetospheric accretion models couple the dipper phenomenon to the stellar rotation and also provide a mechanism for lifting dusty material out of the midplane.

\section{A Unification Paradigm for Young Dippers}

\begin{figure}
\includegraphics[width=3.4in, trim= 0 0 0 0 ]{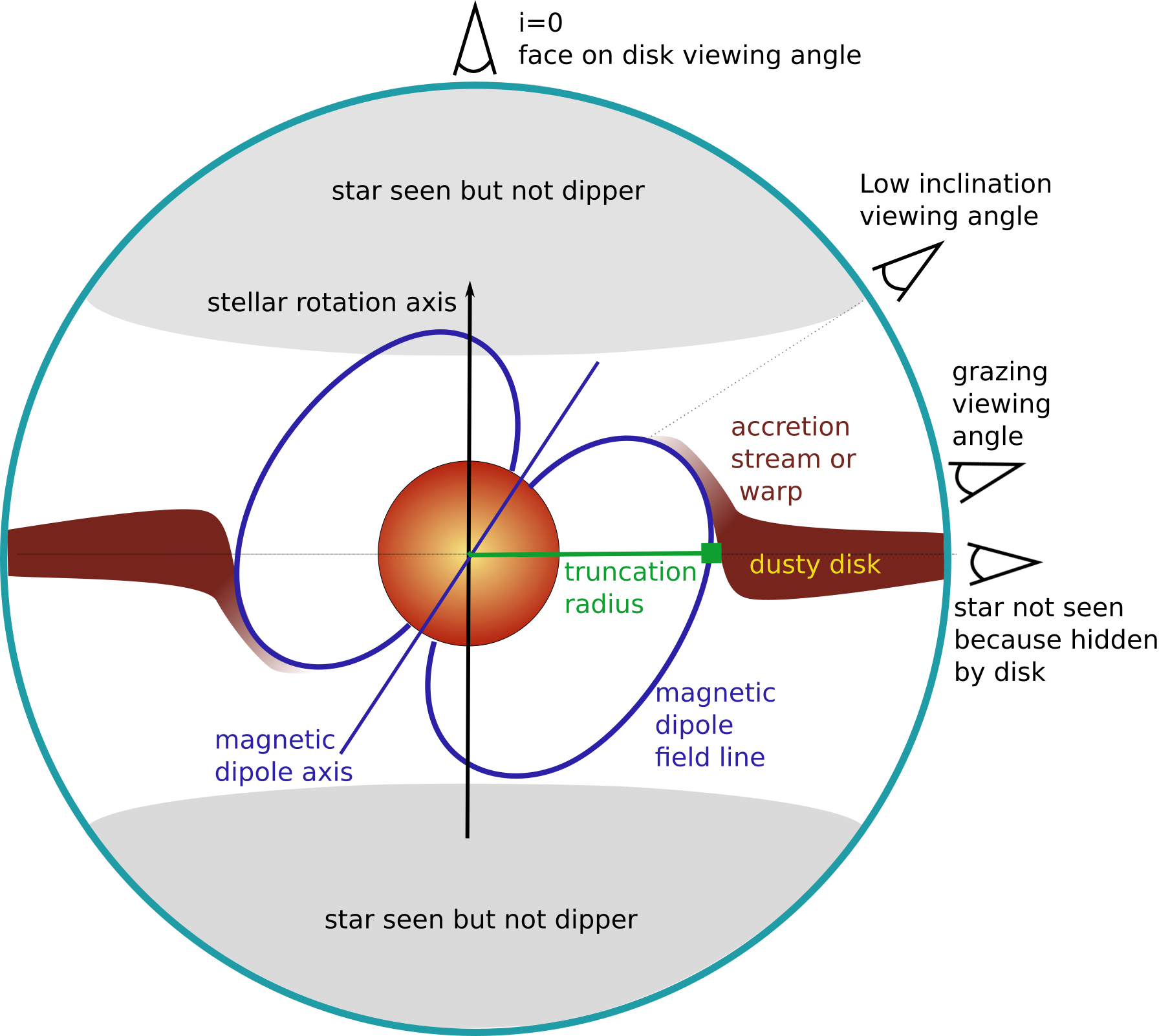}
\caption{Magnetospheric truncation with a misaligned magnetic field. The magnetic pressure causes accreting disk material to pile up at the truncation radius and the increased pressure pushes gas out of the midplane forming a warp or a stream. If the stream is cool enough for dust to survive then a viewer can see a dip in the light curve. Only for low inclination (nearly face-on) viewer orientations would a viewer see an unobstructed view of the star and not see dippers. For a narrow cylinder of orientation angles, the circumstellar disk itself would occult the star, removing it from discovery in optical surveys. The magnetospheric accretion stream model for dippers not only provides a mechanism for lifting gas out of the plane but could account for the large fraction of young stars exhibiting the dipper phenomenon.
}
\label{fig:stream}
\end{figure}

\begin{figure}
\includegraphics[width=3.3in, trim= 0 0 0 0 ]{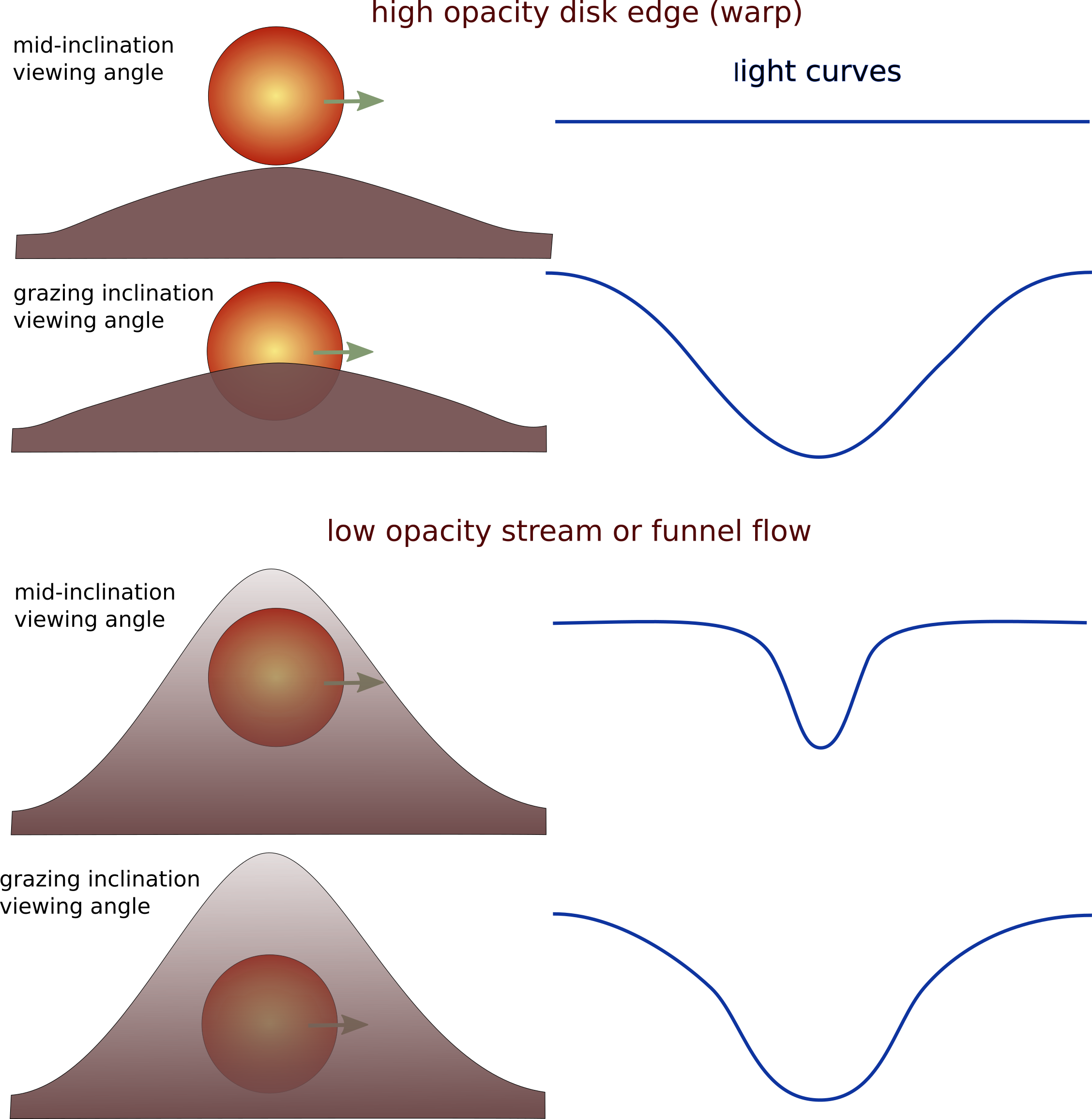}
\caption{As the disk rotates, the accretion stream crosses in front of a star, shown here as seen by a distant viewer at the low and  grazing  viewing angles, as shown in Figure \ref{fig:stream}. The top two panels show an optically-thick warp with associated light curves on the right. The bottom two panels show an optically-thin wide funnel flow or stream with associated light curves on the right. If the dip is due to an optically-thick warped disk edge (top two panels), then a dipper would be seen only for grazing orientations. However, if the stream is optically thin and wide, then a wide dip can be seen over a wide range of orientation angles.
 }
\label{fig:stream_side}
\end{figure}

The unification paradigm for active galactic nuclei (AGN) is the proposal that the diversity of observed AGNs can be explained by a small number of underlying physical parameters, one of which is the orientation angle of the accretion disk with respect to the viewer \citep{netzer15}. For AGN, underlying parameters include the black hole mass and accretion rate. For low-mass accreting young stellar objects, we explore a similar paradigm. As is true for the AGN unification paradigm, the orientation angle with respect to the viewer should affect observed properties of the young stellar object. We assume that the spin axis of the star is equal to that of the inner circumstellar disk. We describe the disk inclination, $i$, by the angle between the line of sight and the spin axis of the disk. High inclination disks ($i = 90^\circ$) are edge-on and low inclination disks ($i=0^\circ$) are face-on.  

The magnetic pressure from the star's dipole field causes material to pile up at the magnetospheric truncation radius and the matter pressure gradient and gravitation forces push gas up or down out of the midplane, forming a warp or two accretion streams that approximately follow magnetic field lines to the star \citep{romanova03, romanova04, romanova08, romanova12, romanova15}. For an illustration, see Figure \ref{fig:stream}. If the stream is cool enough for dust to survive, then an external viewer would see a dip in the light curve when the stream passes in front of the star. 
Models of accretion streams show that much of the stream is well above the dust sublimation limit \citep[e.g.,][]{1994Hartmann,1996Martin,2001Muzerolle}. \citet{2014Petrov} measured the accretion stream temperature of T~Tauri star S~CrA and found a gradient from $\sim$5000-7000 K extending most of the stream. However, most of the stars in this sample are cooler than the one studied by \citet{2014Petrov} and the inner edge is well below the dust sublimation limit, so we expect the dust to survive in the stream near the base and start to sublimate with the rising temperature after the dust is lifted above the midplane.
 If the stream extends well above the disk midplane and the dust survives for most of that height, then the star will present as a dipper for a large fraction of the total solid angle of possible viewer orientations. Only for nearly face-on orientations (low $i$) would a viewer see an unobstructed view of the star and not see dippers. The paradigm could potentially account for the large fraction of young stars exhibiting the dipper phenomenon \citep{2010Alencar,2014Cody}.

The accretion streams cannot be both optically thick and taller and wider than the diameter of the star, otherwise the dip depth would be nearly unity and none of the dips are that deep.
The base of the accretion stream or the warp must be wider than the stellar diameter to account for the length of the dips as, in most stars considered here, the dips are wider than the crossing time; see Figure \ref{fig:slope}. If the occulting material is optically thick, then at most only part of the star can be covered during a dip to match dip depths. This would be consistent with the edge of a warped disk occulting the star during part of the orbit. In contrast, if the occulting material is optically thin, then the whole star could be covered during a dip. These two possibilities were discussed in terms of warped disk and funnel flow models by \citet{stauffer15} in their sections 7.2 and 7.3. In Figure \ref{fig:stream_side}, we illustrate how material occults the star for two different disk orientations for the optical thick warp (top two panels) and for the optically thin wide stream or funnel flow (bottom two panels). This illustration also shows how the light curve might depend on orientation angle.

Ordinarily, one would favour an optically-thick model as fine tuning is required for the opacity to be $\sim$ 1.  However, the optically-thin wide stream model has an advantage when considering the fraction of orientation angles that allow dipper phenomena. If the disk is optically thick, then a dipper with durations longer than a crossing timescale would only be seen at disk-grazing inclinations (taking into account the ratio $R_T/R_\star \sim R_\mathrm{cor}/R_\star \sim 5$). However, if the stream is optically-thin, then a wide range of disk orientation angles (see Figure \ref{fig:stream_side}) would allow the viewer to see moderate optical depth but long (compared to the crossing time) dips in the light curve. It is easier to account for the large fraction of young stars exhibiting dipper phenomena for a unified model with low opacity but wide accretion streams.
%Figure \ref{fig:stream_side} also shows how the light curve shape might depend on orientation angle.

If the stream is wide at the base and narrows as the material flows onto the star, then wide dips would be associated with grazing orientation angles and narrower dips would be associated with lower-inclination viewing angles, as illustrated in Figure \ref{fig:stream}. We searched for and failed to find the correlations between dip depth and width (in units of the crossing timescale) that we expected, since more of the star would be covered by the stream at grazing inclinations. However, if the stream is dense near the top of the stream as suggested in simulations (see Figure 8 by \citealt{romanova13}), then mid-inclination-oriented systems might have dips as deep as grazing-oriented systems. The very top of the stream is low density but a little below is very dense. The top of the stream is also hotter, so less dust would survive there, reducing the dip depth.

%Are the accretion rates consistent with a moderate opacity accretion stream?
The opacity of an accretion stream depends on the accretion rate. A rectangular stream of width $W_s$ (which could be a function of height above the disk midplane) moving at velocity $V_s$ upwards away from the midplane has mass surface density 
\begin{equation}
\Sigma_s = \frac{\dot M}{W_s V_s}.
\end{equation}
We constrain the stream width with the crossing timescale which depends on the diameter of the star (equation \ref{eqn:tcross}). We describe the velocity in units of the circular velocity at the corotation radius, $V_\mathrm{cor}$ (equation \ref{eqn:Vcor}). This gives a mass surface density of 
\begin{eqnarray}
\Sigma_s &\approx  & \frac{\dot M}{2 R_\star} \left( \frac{P}{2\pi} \right)^\frac{1}{3} \left( GM_\star \right)^{-\frac{1}{3}} \left( \frac{W_s}{2 R_\star}\right)^{-1} \left( \frac{V_s}{V_\mathrm{cor}}\right)^{-1} .
\end{eqnarray}
Replacing the ratio $W_s/(2R_\star)$ with a ratio of a dip duration to crossing timescale $t_\mathrm{dip}/t_\mathrm{cross}$ which we can measure from the light curves and evaluating the surface density
\begin{eqnarray}
\Sigma_s &\approx  & 
 0.0064 {\rm ~g~cm}^{-2} ~  \dot M_{-9} \left( \frac{M_\star}{0.5 M_\odot} \right)^{-\frac{1}{3}} \left( \frac{R_\star}{2 R_\odot } \right)^{-1}  \nonumber \\ 
 && \times \left(\frac{P}{3 \ {\rm day}}\right)^\frac{1}{3} \left( \frac{t_\mathrm{dip}/t_\mathrm{cross}}{3} \right)^{-1} \left( \frac{V_s}{V_\mathrm{cor}} \right)^{-1}
\end{eqnarray}
where $\dot M_{-9}$ is the accretion rate in units of $10^{-9} M_\odot {\rm yr}^{-1}$. From this we estimate a hydrogen column depth $N_H \sim \Sigma_s/m_H$ where $m_H$ the mass of a hydrogen atom and using the interstellar value for extinction in I~band
$$(A_I/N_H)_\mathrm{interstellar} \sim 3 \times 10^{-22} {\rm mag~cm}^{-2}$$
\citep{fitzpatrick99},
we estimate the corresponding extinction in $I$ band 
\begin{eqnarray}
A_I &\sim& 1.1 ~  \left( \frac{(A_I/N_H)_\mathrm{YSO}}{(A_I/N_H)_\mathrm{interstellar}} \right)\dot M_{-9}\nonumber \\ 
&& \times \left( \frac{M_\star}{0.5 M_\odot} \right)^{-\frac{1}{3}} \left( \frac{R_\star}{2 R_\odot } \right)^{-1} \left(\frac{P}{3 \ {\rm day}}\right)^\frac{1}{3} 
\nonumber \\ 
&& \times \left( \frac{t_\mathrm{dip}/t_\mathrm{cross}}{3} \right)^{-1}
%\left( \frac{W_s/(2 R_\star)}{3} \right)^{-1}
\left( \frac{V_s/V_\mathrm{cor}}{1} \right)^{-1}.
\end{eqnarray}
I~band ($0.8 \mu$m) is chosen because the stars are red and the \textit{Kepler} mission bandpass (0.4 to 0.9 $\mu$m) is wide. A lower velocity in the stream itself would increase the column depth during a dip, whereas a lower dust to gas ratio would decrease the extinction. It is interesting to notice that intermediate levels of opacity (near 1) are expected. Since optically-thin funnel flows or stream can resolve the problem of how to account for the large number of stars displaying dipper phenomena, and low levels of opacity are implied by the low accretion rates, the low-opacity stream model provides an attractive possibility for a unified model.

One consequence of our illustration in Figure \ref{fig:stream} is that the accretion stream or warped inner disk would cast shadows on the outer disk.  A face-on disk might not exhibit dipper phenomena but high-resolution images of the outer disk might exhibit shadows of the inner disk.  If the inner disk nearly corotates with the star, then we would expect the shadows to rotate at this period. The outer disk of MP~Muscae, as seen by high-resolution imaging \citep{2016Wolff,schneider14}, exhibits 3 month (or shorter) temporal variability in azimuthal asymmetry. This star is a pre-main-sequence K1 star with a 5 day rotation period, cool enough and with a long enough period that it could be a dipper if it were oriented at a higher inclination. Future study of this and other objects could test this consequence of a unification paradigm for dusty disk edges in young stellar objects.

\subsection{Magnetic dipole tilt angle and current accretion rate} \label{sec:magacc}

\begin{figure}
\includegraphics[width=3.4in, trim= 0 0 0 0 ]{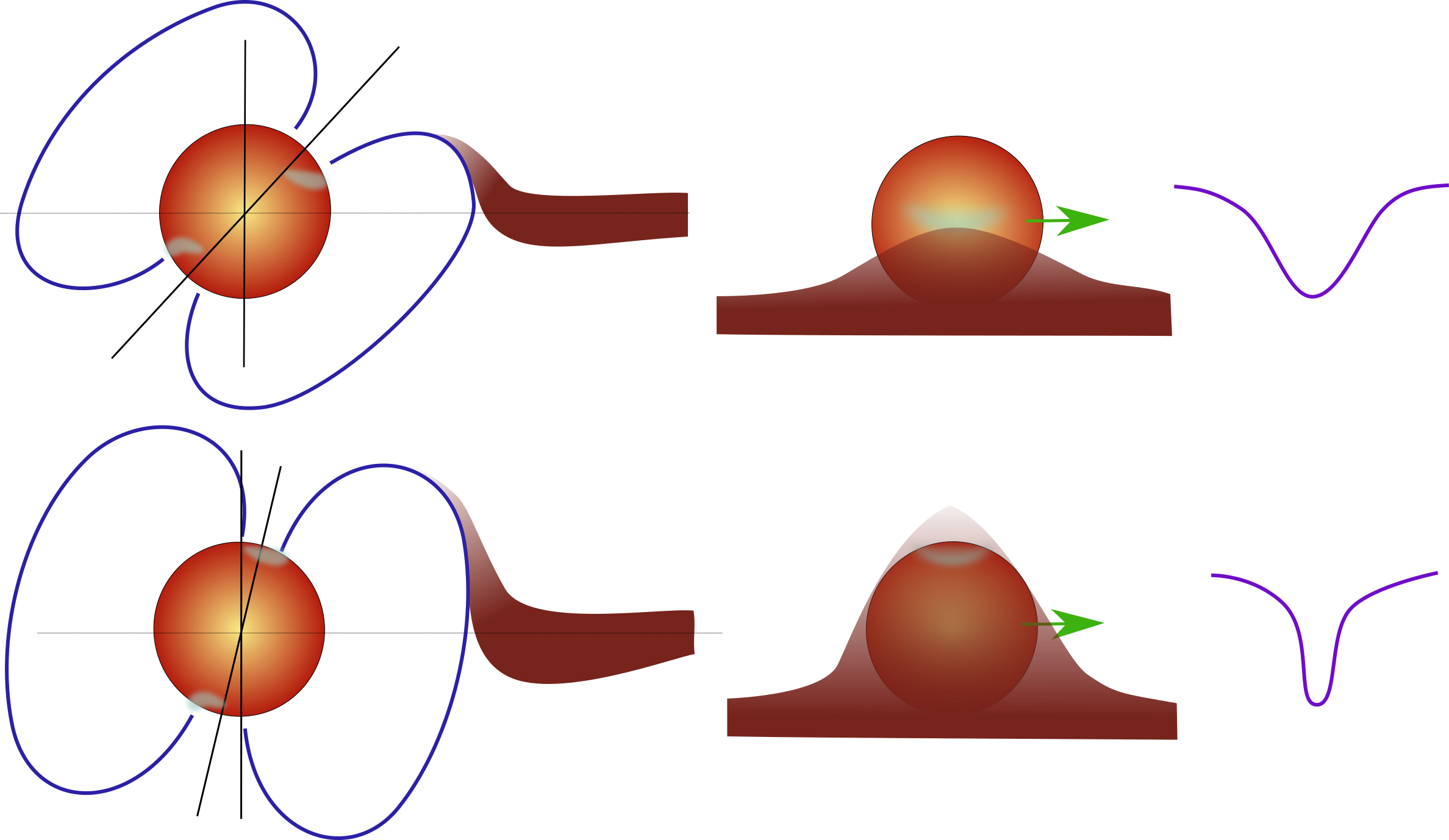}
\caption{On the left we illustrate side views of accretion streams for a low-inclination (top) dipole magnetic field tilt angle (measured with respect to the stellar spin axis) and a higher tilt angle (bottom). The dusty accretion streams must be lifted higher above the disk midplane to accrete onto the star when the tilt axis is lower.  This implies that low-tilt-angle systems would have a larger solid angle of viewing orientation for dippers to be seen than would high-tilt-angle systems. Low-tilt-angle systems would have nearly-polar hotspots and so lower-amplitude modulations caused by hotspots in the light curve.
 }
\label{fig:stream_tilt}
\end{figure}

A key parameter affecting magnetospheric accretion flow geometry and stability is the tilt axis of the star's magnetic dipole \citep{romanova15, 2016Blinova}. 
 \citet{2015McGinnis} associated aperiodic dippers in NGC~2264 with these unstable accretion ``tongues'' and periodic dippers with stable accretion funnels.
If the dipole axis is aligned with the star's spin axis, then the disk edge is more prone to the magnetic Rayleigh-Taylor instability which enables matter to penetrate the magnetosphere via several chaotically-formed ``tongues'' or ``fingers''  \citep{romanova08, 2016Blinova}. 
These ``tongues'' can reach out of the midplane and occult the star, resulting in an aperiodic dipper for mid-high to grazing viewing inclinations. Stable accretion can also occur for stars with a dipole axis nearly aligned with the spin axis \citep{2016Blinova} and form broad funnel flows which result in periodic dippers for a larger range of viewing orientations.
If the dipole axis is near the circumstellar disk midplane, then gas can fall directly onto the star without leaving the midplane. We expect streams to cover a larger solid angle of viewing orientation where they would be seen crossing the star for lower magnetic dipole tilt angles as measured from the star's rotation axis. Hence there may be a bias toward detecting lower-tilt-angle systems, especially for periodic dips, as a larger fraction of orientation angles would present as dippers (see Figure \ref{fig:stream_tilt} for an illustration). 
We expect the deeper and wider dips to be periodic since stable accretion funnels are larger than the unstable ``tongues''.  However, we have searched for and failed to find correlations between dip depth or width and periodicity of the dips.
 
 As discussed by \citet{romanova04, kulkarni13, romanova15}, there may be a relation between hotspot latitude (as viewed by the observer) and phase (with respect to the maximum stream density) and the magnetic tilt axis. We have illustrated this crudely in Figure \ref{fig:stream_tilt}. If multicolour photometry and spectroscopy could differentiate between stream and hot spot location, perhaps it would be possible to measure the dipole tilt axis. 
 Variability from dust extinction and hotspots have different wavelength dependencies so that contribution from each source can be differentiated \citep{2015Venuti}. \citet{romanova04} showed that the shape of the hotspot depends on the tilt of the magnetic dipole axis and the different hotspot shape causes small differences in the light curve but has degeneracies with viewing inclination. The stream can also occult the hotspot, which would constrain the viewing orientation.
  So while the shape of the dip in a single band might not place constraints on the tilt axis, additional colour information might break the degeneracies between the dipole tilt angle and the viewing angle.

Even though we suspect that the magnetospheric truncation radii are near the corotation radius, short timescale variations in accretion rate may change the current truncation radius, $R_T$, with a higher accretion rate decreasing $R_T$.
% AA~Tau itself lost its periodicity in 2011 \citep{bouvier03, bouvier13}.  
When comparing known variable stars in NGC~2264 between 2008 and 2011, \citet{sousa16} found that $\sim30\%$ of the stars changed their light curve morphology, mostly AA~Tau-like to aperiodic or aperiodic to AA~Tau-like, as summarized in their Table 5. 
In their sample of extinction dominated dippers, \citet{2015McGinnis} found about half of the stars changed their light curve morphology between aperiodic and periodic.
 The stability of the accretion flow strongly depends on the ratio of $R_T/R_\mathrm{cor}$, called the fastness parameter. Lower values of the fastness parameter which corresponds to higher accretion rates are more likely to be unstable (see Figure 1 by \citealt{2016Blinova}). One hypothesis is that periods of high accretion are associated with more instability. We might be seeing evidence of this in some of the light curves, for example EPIC~204489514 may show stronger star spot variation during times when there are deeper dips. Using multiwavelength observations to differentiate between star spots and extinction, one might look for evidence of lower-latitude hot spots from the lower-latitude unstable ``tongues'' during epochs of higher accretion rates.

%The magnetic dipole tilt axis could affect the likelihood of finding a dipper. If the tilt angle is low, then streams must be pushed further out of the disk midplane and a larger fraction of orientation angles should present as dippers. Previous studies have predicted that the magnetic dipole  axis tilt and current disk accretion rate would affect the stability of the accretion flow \citep{romanova15}. Hot spot latitude might also be sensitive to the dipole  axis tilt and current accretion rate. Transitions between quasi-periodic and aperiodic light curves might be manifestations of variation in accretion rate.

\section{Summary}

In this paper, we have reexamined the properties of the young star dippers from Upper Sco and $\rho$ Oph discovered by \citet{2016A}.
Young stars rotate and are magnetically-active resulting in star spots which can cause quasi-periodic variations in the light curve. Rotation periods measured from the light curves of our sample are typical of stellar rotation periods of young low-mass stars. However, the spectrograms computed for most of the the light curves show a single dominant period that is also associated with dips in the light curve, and do not show a different period even during times when the light curve lacks dipper events. The single dominant period we associate with both stellar rotation and the period of quasi-periodic dips indicates the dust obscuring the star is corotating with the star, confirming previous work by \citet{stauffer15}. 
We also find that maximum light curve slopes of both aperiodic and periodic dippers are correlated (0.78) with those computed for optically-thick blocks corotating with the star.  Aperiodic dippers do not have a period that can be compared to the star's rotation period but the shape of the dips is consistent with material near corotation obscuring the star. %1

We computed the effective temperature in the disk at the radius associated with this rotation period and find that it is below the dust sublimation temperature. This confirms that the corotation region of the disk is capable of obscuring the star with dusty material.  Using stellar tracks, we show that only cool stars are likely to have corotation radii that are below the dust sublimation temperature. If dippers are associated with corotating dust, then they are most likely to be found among low-mass stars, which explains the preference of the dippers discovered by \citet{2016A} to be low-mass stars.  Furthermore, we expect a correlation between period and stellar effective temperature since dust must be further from a hotter star to survive, consistent with the correlation found by \citet{2016A}. 
UX~Orionis variable-type stars are an exception since they are hot and fast rotators, causing their corotation radius to be within the dust sublimation radius. As the stellar mass increases, a different mechanism is required to produce dippers. %2

Using roughly-estimated dipole surface magnetic fields and accretion rates, we find that magnetospheric truncation radii could be near our estimated corotation radii within our large uncertainties. 
Magnetospheric truncation radii should not differ significantly from the corotation radii. If the truncation radius is large compared to the corotation radius, then the system should not accrete, as it would be in the propeller regime, and if much smaller compared to the corotation radius, the disk edge would be too hot for dust to survive.
 Currently, few magnetic field measurements have been made for dippers and accretion rates are highly uncertain.
Better measurement of magnetic field and accretion rates for the dippers might confirm the near match between corotation and disk truncation radii. If magnetospheric disk truncation radii are approximately at corotation, then it would confirm predictions of strong star-disk interaction (e.g., \citealt{konigl91}), as this would imply that stellar rotation is controlled by disk accretion. %4

An advantage of magnetospheric truncation models is that they provide a mechanism for lifting material out of the disk midplane so dust occults the star over a large range of viewing angles and links the dipper phenomenon to the stellar rotation. We find that magnetospheric truncation radii are only below the dust sublimation temperature if the accretion rates are low, $\dot{M}\lesssim 10^{-8}$. The association of truncation radii with the material causing the dips implies that only low accretion rate objects will be dippers. The preference of low-mass and low-accretion-rate stars to exhibit dippers is consistent with those found in the Upper Sco and $\rho$ Oph star formation regions by \citet{2016A} and the dippers of NGC~2264 \citep{2015Venuti,sousa16}. Many of the stars in the sample are classified as WTTS which are typically no longer accreting but \citet{2016A} found weak accretion signatures for most of the 10 stars in the sample and found IR excess consistent with disks capable of supporting accretion for all the stars. The stars in our sample are most likely near the CTTS-WTTS transition with moderately-evolved disks and weak accretion so magnetospheric accretion model applies.
%3 

%Unification of active galactic nuclei (AGN) is the proposal that the diversity of observed AGNs can be explained by a small number of underlying physical parameters, one of which is the orientation angle of the viewer.
 Building upon previous studies \citep{bouvier13, stauffer15, 2015McGinnis, 2016A}, we explore a unification paradigm for dippers. If accretion streams have visible band opacity less than but near 1, then we can simultaneously account for the approximate depth and duration of the dips (a few stellar diameter crossing times) and explain why objects would exhibit dipper phenomena for a large fraction of possible disk orientation angles, potentially accounting for the large fraction of young stars that are dippers. 
The orientation can also account for the difference in dipper morphology with the narrow streams being associated with more mid-inclination viewing angles and wide AA~Tau like dips occurring for near grazing orientations. Other underlying parameters such as accretion rate and dipole tilt angle would also change the shape of the dips and affect the stability of the accretion streams. Unstable accretion streams would result in aperiodic dippers whereas stable accretion results in periodic behaviour. %5

Multicolour light curves can be used to differentiate between star spots and extinction \citep{2015Venuti} and test our hypothesis that the stream opacities are near 1 in visible bands. Multiwavelength studies might also help break the degeneracy between viewing orientation and dipole tilt angle by possible constraining hot spot latitude. Spectroscopic studies might measure the velocity in different parts of the accretion flow and probe the temperature and composition of the flow. 
\citet{2014Petrov} probed the temperature and composition for hot regions of the stream far from the base. Dippers provide an opportunity to probe the base of the stream. 
As dippers are associated with somewhat evolved circumstellar disks, the composition of the flow might have been modified by planet formation in the disk. Simulations could be used to predict the statistics of light curve properties taking into account a random sampling of disk orientation angles. %7

Clumps or vortices in the disk have been proposed as alternative explanations for dippers \citep{2014Fu, 2015Crn}. However, these mechanisms don't explain why the dipper periods are similar to stellar rotation periods, the preference for many low-mass, low-accretion-rate stars to display dipper phenomena or give a clear associated physical mechanism that lifts material out of the midplane to account for the large fraction of T~Tauri stars displaying dips. 
The UX~Orionis variable star HD~142666 is an exception to the obscuring dust near the corotation radius and some of the aperiodic stars in our sample could also be exceptions. More dippers are being identified in Upper Sco and $\rho$ Oph that may be exceptions as well. These stars require a different mechanism but are only a small fraction of dippers. 
The cold dusty magnetospheric accretion model for dippers can be tested statistically with surveys and with more detailed models and measurements of individual objects. %However, we may in future identify phenomena that cannot be explained by magnetospheric truncation models and might require alternative mechanisms for explanation. %8

\section*{Acknowledgements}
Some of the data presented in this paper were obtained from the Mikulski Archive for Space Telescopes (MAST). STScI is operated by the Association of Universities for Research in Astronomy, Inc., under NASA contract NAS5-26555. Support for MAST for non-HST data is provided by the NASA Office of Space Science via grant NNX09AF08G and by other grants and contracts.
This paper includes data collected by the Kepler mission. Funding for the Kepler mission is provided by the NASA Science Mission directorate.

E.~H.~B. and A.~Q. was supported in part by NASA grant NNX13AI27G. 
E.~H.~B.'s research was supported by an appointment to the NASA Postdoctoral Program for the Nexus for Exoplanet System Science, administered by Universities Space Research Association under contract with NASA.
J.~K.'s research on X-ray-active nearby young stars is supported in part by NASA Astrophysics Data Analysis Program grant NNX16AG13G to RIT.
We thank the following for helpful discussion and correspondence:  Dan Watson, Shane Fogerty, Fred Moolekamp, Mark Wyatt, and Eddie Hanson.

\end{document}